\documentclass[usenatbib]{mnras}
\usepackage{amsmath}
\usepackage{natbib}
\usepackage{epsfig}
\usepackage{graphicx}
\usepackage{amssymb}
\usepackage{url}

\def\gtsima{$\; \buildrel > \over \sim \;$}
\def\ltsima{$\; \buildrel < \over \sim \;$}
\def\gtrsim{\lower.5ex\hbox{\gtsima}}
\def\lesssim{\lower.5ex\hbox{\ltsima}}

\newcommand{\msun}{$\mathrm{M}_{\odot}$}

\usepackage{amsmath}	
\usepackage{amssymb}	
\usepackage{mathrsfs}

\begin{document}

\title[Merging black holes in young star clusters]{Merging black holes in young star clusters}
\author[Ugo Niccol\`o Di Carlo et al.]{ \parbox{\linewidth}{Ugo N. Di Carlo$^{1,2,3}$\thanks{E-mail: ugo.dicarlo@inaf.it}, Nicola Giacobbo$^{2,3,4}$, Michela Mapelli$^{2,3,4,5}$\thanks{E-mail: michela.mapelli@unipd.it}, Mario Pasquato$^{2,3}$,  Mario Spera$^{2,3,4,5,6,7}$, Long Wang$^{8,9}$, Francesco Haardt$^{1}$}
\vspace{1cm}
\\
$^{1}$Dipartimento di Scienza e Alta Tecnologia, University of Insubria, Via Valleggio 11, I--22100, Como, Italy
\\
$^{2}$INAF-Osservatorio Astronomico di Padova, Vicolo dell'Osservatorio 5, I--35122, Padova, Italy
\\
$^{3}$INFN, Sezione di Padova, Via Marzolo 8, I--35131, Padova, Italy
\\
$^{4}$Dipartimento di Fisica e Astronomia `G. Galilei', University of Padova, Vicolo dell'Osservatorio 3, I--35122, Padova, Italy
\\
$^{5}$Institut f\"ur  Astro- und Teilchenphysik, Universit\"at Innsbruck, Technikerstrasse 25/8, A--6020, Innsbruck, Austria
\\
$^{6}$Center for Interdisciplinary Exploration and Research in Astrophysics (CIERA), Evanston, IL 60208, USA
\\
$^{7}$Department of Physics \& Astronomy, Northwestern University, Evanston, IL 60208, USA
\\
$^{8}$Argelander-Institut f\"ur Astronomie, Auf dem H\"ugel 71, 53121, Bonn, Germany
\\
$^{9}$RIKEN Advanced Institute for Computational Science, 7-1-26 Minatojima-minami-machi, Chuo-ku, Kobe, Hyogo 650-0047, Japan
}
\maketitle \vspace {7cm}
\bibliographystyle{mnras}

\begin{abstract}
  Searching for distinctive signatures, which characterize different formation channels of binary black holes (BBHs), is a crucial step towards the interpretation of current and future gravitational wave detections. 
  Here, we investigate the demography of merging BBHs in young star clusters (SCs), which are the nursery of massive stars. We performed $4\times{} 10^3$ N-body simulations of SCs with metallicity $Z=0.002$, initial binary fraction $0.4$ and fractal initial conditions,  to mimic the clumpiness of star forming regions. Our simulations include a novel population-synthesis approach based on the code \textsc{MOBSE}.
  We find that SC dynamics does not affect the merger rate significantly, but leaves a strong fingerprint on  the properties of merging BBHs. More than 50 \% of merging BBHs in young SCs form by dynamical exchanges in the first few Myr. Dynamically formed merging BBHs are significantly heavier than merging BBHs in isolated binaries: merging BBHs with total mass up to $\sim{}120$ M$_\odot$ form in young SCs, while the maximum total mass of merging BBHs in isolated binaries with the same metallicity is only $\sim{}70$ M$_\odot$. Merging BBHs born via dynamical exchanges tend to have smaller mass ratios than BBHs in isolated binaries. Furthermore, SC dynamics speeds up the merger: the delay time between star formation and coalescence is significantly shorter in young SCs. In our simulations, massive systems such as GW170729 form only via dynamical exchanges. Finally $\sim{}2$ \% of merging BBHs in young SCs have mass in the pair-instability mass gap ($\sim{}60-120$ M$_\odot$). This represents a unique fingerprint of merging BBHs in SCs. 
\end{abstract}


\maketitle

%

\section{Introduction}
The recent detection of gravitational waves (GWs, \citealt{abbottGW150914}) by LIGO \citep{LIGOdetector} and Virgo \citep{VIRGOdetector} has opened up a new way to investigate the Universe. Out of the 11 GW events reported so far, ten have been interpreted as the merger of two black holes (BHs, \citealt{abbottGW150914,abbottGW151226,abbottO1,abbottGW170104,abbottGW170608,abbottGW170814,abbottO2,abbottO2popandrate}) and one as the merger of two neutron stars (NSs, \citealt{abbottGW170817}). The double neutron star (DNS) merger was accompanied by electromagnetic emission almost in all the electromagnetic spectrum, from radio to gamma rays \citep{abbottmultimessenger,abbottGRB,goldstein2017,savchenko2017,margutti2017,coulter2017,soares-santos2017,chornock2017,cowperthwaite2017,nicholl2017,pian2017,alexander2017}.

Seven out of ten binary BH (BBH) mergers observed by the LIGO-Virgo collaboration harbour BHs with mass $\gtrsim{}30$ M$_\odot$, significantly higher than the range inferred from dynamical measurements of BH masses in nearby X-ray binaries \citep{orosz2003,ozel2010,farr2011}. Such massive stellar BHs are expected to form from the direct collapse of massive metal-poor stars (e.g. \citealt{heger2003,mapelli2009,mapelli2010,belczynski2010,fryer2012,mapelli2013,spera2015,spera2017}). Dynamical interactions in dense environments are also expected to significantly affect the mass spectrum of BHs (e.g. \citealt{portegieszwart2000,portegieszwart2002,portegieszwart2004,gurkan2006,giersz2015,mapelli2016}). Alternatively, primordial BHs, which were predicted to form from gravitational instabilities in the very early Universe \citep{carr2016,raccanelli2016,sasaki2016,scelfo2018} might also have a mass of $\sim{}30-40$ M$_\odot$.

Not only the mass spectrum of compact objects but also the formation channels of compact-object binaries are matter of debate. Double compact objects might result from the evolution of isolated stellar binaries, i.e. systems of two stars which are gravitationally bound since their birth. An isolated binary is expected to undergo a number of processes during its life (including mass transfer, common envelope episodes, tidal forces and supernova explosions). Binary population-synthesis codes are generally used to describe the evolution of isolated binaries and to predict whether they give birth to double compact objects (e.g. \citealt{tutukov1973,flannery1975, bethe1998, portegieszwart1998,  portegieszwart2000,belczynski2002,voss2003, podsiadlowski2004,podsiadlowski2005,belczynski2007,  belczynski2008,bogomazov2007,  moody2009,dominik2012, dominik2013,  dominik2015,  mapelli2013, mennekens2014,tauris2015, tauris2017, demink2015,demink2016,marchant2016,belczynski2016,chruslinska2018,mapelli2017,giacobbo2018,giacobbo2018b,giacobbo2018c,mapelli2018,mapelli2018b,kruckow2018,shao2018,spera2018}).

Alternatively, dynamics in dense environments (e.g. young star clusters, open clusters, globular clusters and nuclear star clusters) can drive the formation and the evolution of compact-object binaries (e.g. \citealt{sigurdsson1993,kulkarni1993,sigurdsson1995,portegieszwart2000,colpi2003,oleary2006,sadowski2008,oleary2009,downing2010,downing2011,mapelli2013,mapelli2014,ziosi2014,rodriguez2015,rodriguez2016,antonini2016,mapelli2016,kimpson2016,hurley2016,oleary2016,askar2017,askar2018,zevin2017,samsing2018,rodriguez2018,antonini2018,rastello2018}). For example, dynamical exchanges can bring BHs into existing stellar binaries. BHs are particularly efficient in acquiring companions via exchanges, because they are among the most massive objects in a star cluster (SC) and exchanges tend to produce more and more massive binaries \citep{hills1980}. For this reason, \cite{ziosi2014} find that $\gtrsim{}90$ per cent of BBHs in open clusters form via dynamical exchanges. Binaries formed through exchanges have a characteristic signature: they tend to be more massive than average, have high initial orbital eccentricity and mostly misaligned spins.

Furthermore, dynamical hardening via three- or multi-body encounters can also affect the evolution of ``hard'' binaries  (i.e. binaries whose binding energy is higher than the average kinetic energy of other SC members \citealt{heggie1975}), by shrinking their orbital separation until they enter the regime where GW emission proceeds efficiently (see e.g. \citealt{mapelli2018review} for a recent review). Finally, ``soft'' binaries (i.e. binaries whose binding energy is lower than the average kinetic energy of other SC members) might even be broken by three-body and multi-body encounters. Other dynamical processes which might affect the evolution of compact-object binaries include runaway collisions (e.g. \citealt{mapelli2016}), Spitzer's instability \citep{spitzer1969}, Kozai-Lidov resonances \citep{kozai1962,lidov1962}, and dynamical ejections (e.g. \citealt{downing2011}).

Dynamics is a crucial ingredient to understand the formation of compact-object binaries, because massive stars (which are thought to be the progenitors of BHs and NSs) form preferentially in young SCs and associations \citep{lada2003,portegieszwart2010}. Thus, it is reasonable to expect that compact objects and their stellar progenitors participate in the dynamics of their parent SC, before being ejected or scattered into the field.

Despite this, most studies of BH dynamics neglect young SCs and star forming regions, with few exceptions \citep{portegieszwart2002,banerjee2010,ziosi2014,goswami2014,mapelli2016,banerjee2017,banerjee2018,fujii2017,rastello2018}. The relative scarcity of studies about BHs in young SCs is partially due to the fact that these are small, generally clumpy and asymmetric stellar systems, mostly composed of few hundred to several thousand stars: they cannot be modelled with fast Monte Carlo techniques, but require expensive direct N-body simulations. Moreover, stellar evolution is a key ingredient in the life of young SCs, because their age is comparable with the lifetime of massive stars: mass loss by stellar winds and supernova (SN) explosions contribute significantly to the dynamical evolution of young stellar systems (e.g. \citealt{mapellibressan2013}). This implies that dynamical models of young SCs should also include stellar evolution through accurate population synthesis.

Furthermore, observations suggest that young embedded SCs are characterised by clumpiness  (e.g. \citealt{cartwright2004,gutermuth2005}) and high fractions of binaries (e.g. \citealt{sana2012}), whereas most simulations of young SCs adopt idealized initial conditions, consisting of monolithic King  models \citep{king1966} and assuming a very small fraction of binaries ($f\sim{}0-0.1$). 

Our aim is to study the demography of double compact objects in young SCs, following a novel approach: we have run a large set of N-body simulations of young SCs with fractal initial conditions (which mimic the clumpy and asymmetric structure of star forming regions, e.g. \citealt{goodwin2004}) and with a high initial binary fraction ($f_{\rm bin}=0.4$). The initial masses of our SCs have been randomly drawn according to a power-law distribution $dN/dM\propto{}M^{-2}$ (from  $M=10^3$ \msun{} to $3\times{}10^4$ \msun{}), consistent with the mass distribution of young SCs in the Milky Way \citep{elmegreen1997,lada2003}. While these SCs host less stars than globular clusters and other massive clusters, they make up the vast majority of the SCs in the Universe \citep{kroupa2002}, and their cumulative contribution to BH statistics may thus be significant. We evolve each SC with an accurate treatment of dynamics \citep{wang2015} and with up to date binary population-synthesis models \citep{giacobbo2018}.

\section{Methods}

The simulations were done using the direct summation N-Body code \textsc{NBODY6++GPU} \citep{wang2015} coupled with the new population synthesis code \textsc{MOBSE} \citep{giacobbo2018}. 

\subsection{Direct N-Body}

\textsc{NBODY6++GPU} is the GPU parallel version of \textsc{NBODY6} \citep{aarseth2003}. It implements a 4th-order Hermite integrator, individual block time–steps \citep{makino1992} and Kustaanheimo-Stiefel (KS) regularization of close encounters and few-body subsystems \citep{stiefel1965}.

A neighbour scheme \citep{nitadori2012} is used to compute the force contributions at short time intervals (\textit{irregular} force/timesteps), while at longer time intervals (\textit{regular} force/timesteps) all the members in the system contribute to the force evaluation. The irregular forces are evaluated using CPUs, while the regular forces are computed on GPUs using the CUDA architecture.

\subsection{Population synthesis}
In its original version, \textsc{NBODY6++GPU} is coupled with the population synthesis code \textsc{BSE} \citep{hurley2000,hurley2002}.  For this work, we modified \textsc{NBODY6++GPU}, coupling it with the new population synthesis code \textsc{MOBSE} \citep{mapelli2017,giacobbo2018,giacobbo2018b,giacobbo2018c,mapelli2018}, an updated version of \textsc{BSE}. \textsc{NBODY6++GPU} and \textsc{MOBSE} are perfectly integrated: they update stellar parameters and trajectories simultaneously during the computation.

\textsc{MOBSE} implements some of the most recent stellar wind models for massive hot stars \citep{vink2001,graefener2008,vink2011,vink2016}, including the impact of the Eddington factor $\Gamma_e$ on mass loss \citep{graefener2008,chen2015}. In \textsc{MOBSE} the mass loss of massive hot stars (O- and B-type main sequence stars, luminous blue variable stars and Wolf-Rayet stars) scales as $\dot{M}\propto Z^\beta$, where $\beta$ is defined as \citep{giacobbo2018b}
\begin{equation}
\beta = \begin{cases}
0.85 & \mathrm{if} \quad \Gamma_e < 2/3\\
2.45-2.4\Gamma_e & \mathrm{if} \quad 2/3 \leq \Gamma_e < 1 \\
0.05 & \mathrm{if}\quad \Gamma_e\geq 1
\end{cases}
\end{equation}

\textsc{MOBSE} includes two different prescriptions for core-collapse supernovae (SNe) from \cite{fryer2012}: the \textit{rapid} and the \textit{delayed} SN models. The former model assumes that the SN explosion only occurs if it is launched $\lesssim{}250$ msec after the bounce, while the latter has a longer timescale ($\gtrsim{}500$ msec). In both models, a star is assumed to directly collapse into a BH if its final Carbon-Oxygen mass is $\gtrsim{}11$ M$_\odot$. For the simulations described in this paper we adopt the \textit{delayed} model. Recipes for electron-capture SNe are also included in MOBSE \citep{hurley2000,fryer2012,giacobbo2018c}.

Prescriptions for pair instability SNe (PISNe) and pulsational pair instability SNe (PPISNe) are implemented using the fitting formulas by \cite{spera2017}. In particular, stars which grow a Helium core mass $64\le{}m_{\rm He}/{\rm M}_\odot\le{}135$ are completely disrupted by pair instability and leave no compact object, while stars with  $32\le{}m_{\rm He}/{\rm M}_\odot<64$ undergo a set of pulsations (PPISN), which enhance mass loss and cause the final remnant mass to be significantly smaller than they would be if we had accounted only for core-collapse SNe. According to our current knowledge of PISNe and PPISNe and to the stellar evolution prescriptions implemented in \textsc{MOBSE}, we expect no compact objects in the mass range $\sim{}60-120$ M$_\odot$ from single stellar evolution. Binary evolution might affect this range. For example, mass accretion onto an evolved star (or the merger between a post main sequence star and a main sequence star) might increase the mass of the Hydrogen envelope without significantly affecting the Helium core: at the time of collapse, such star will have a Helium core mass below the PPISN/PISN range, but a significantly larger Hydrogen envelope. By direct collapse, this star might produce a BH with mass $\ge{}65$ M$_\odot$. Clearly, the possibility of forming BHs with mass in the PPISN/PISN gap depends on the assumptions about efficiency of mass accretion, about mass loss after stellar mergers and about core-collapse SNe.

Thanks to these assumptions for massive star evolution and SNe, the BH mass spectrum predicted by \textsc{MOBSE} depends on progenitor's metallicity (the maximum BH mass being higher at lower metallicity) and is consistent with LIGO-Virgo detections \citep{abbottO1,abbottGW150914,abbottGW151226,abbottGW170104,abbottGW170608,abbottGW170814,abbottO2}.

Natal kicks are randomly drawn from a Maxwellian velocity distribution. A one-dimensional root mean square velocity $\sigma_{\rm CCSN}=265$ km s$^{-1}$ and $\sigma{}_{\rm ECSN}=15$ km s$^{-1}$ are adopted for core-collapse SNe \citep{hobbs2005} and for electron-capture SNe \citep{dessart2006,jones2013,schwab2015,giacobbo2018c}, respectively. Kick velocities of BHs are reduced by the amount of fallback as $V_{\mathrm{KICK}}=(1-f_{\mathrm{fb}})\,{}V$, where $f_{\mathrm{fb}}$ is the fallback parameter described in \cite{fryer2012} and $V$ is the velocity drawn from the Maxwellian distribution.

Binary evolution processes such as tidal evolution, Roche lobe overflow, common envelope and gravitational-wave energy loss are taken into account as described in \cite{hurley2002}. In particular, the treatment of common envelope is described by the usual two parameters, $\alpha{}$ and $\lambda{}$. In this work, we assume $\alpha{}=3$, while $\lambda{}$ is derived by \textsc{MOBSE} as described in \cite{claeys2014}.

Orbital decay and circularisation by gravitational-wave emission are calculated according to \cite{peters1964} without explicitly including post-Newtonian terms. The standard version of \textsc{BSE} calculates gravitational-wave energy loss only if the two binary members are closer than 10 R$_{\odot}$, which has been shown to lead to a serious underestimate of the merger rate of eccentric binaries in dynamical simulations \citep{samsing2018}. In \textsc{MOBSE}, we remove the 10 R$_{\odot}$ limit: gravitational-wave decay is calculated for all binaries of two compact objects (white dwarfs, neutron stars and black holes).

Finally, if two stars merge during the N-body simulations, the amount of mass loss is decided by \textsc{MOBSE}, which adopts the same prescriptions as \textsc{BSE}, but for one exception: if a star merges with a BH or a neutron star, \textsc{MOBSE} assumes that the entire mass of the star is immediately lost by the system and the BH (or neutron star) does not accrete it (while the version of \textsc{BSE} implemented in \textsc{NBODY6++GPU} assumes that the entire mass is absorbed by the compact object).

\subsection{Initial conditions}
\begin{figure}
  \center{
    \epsfig{figure=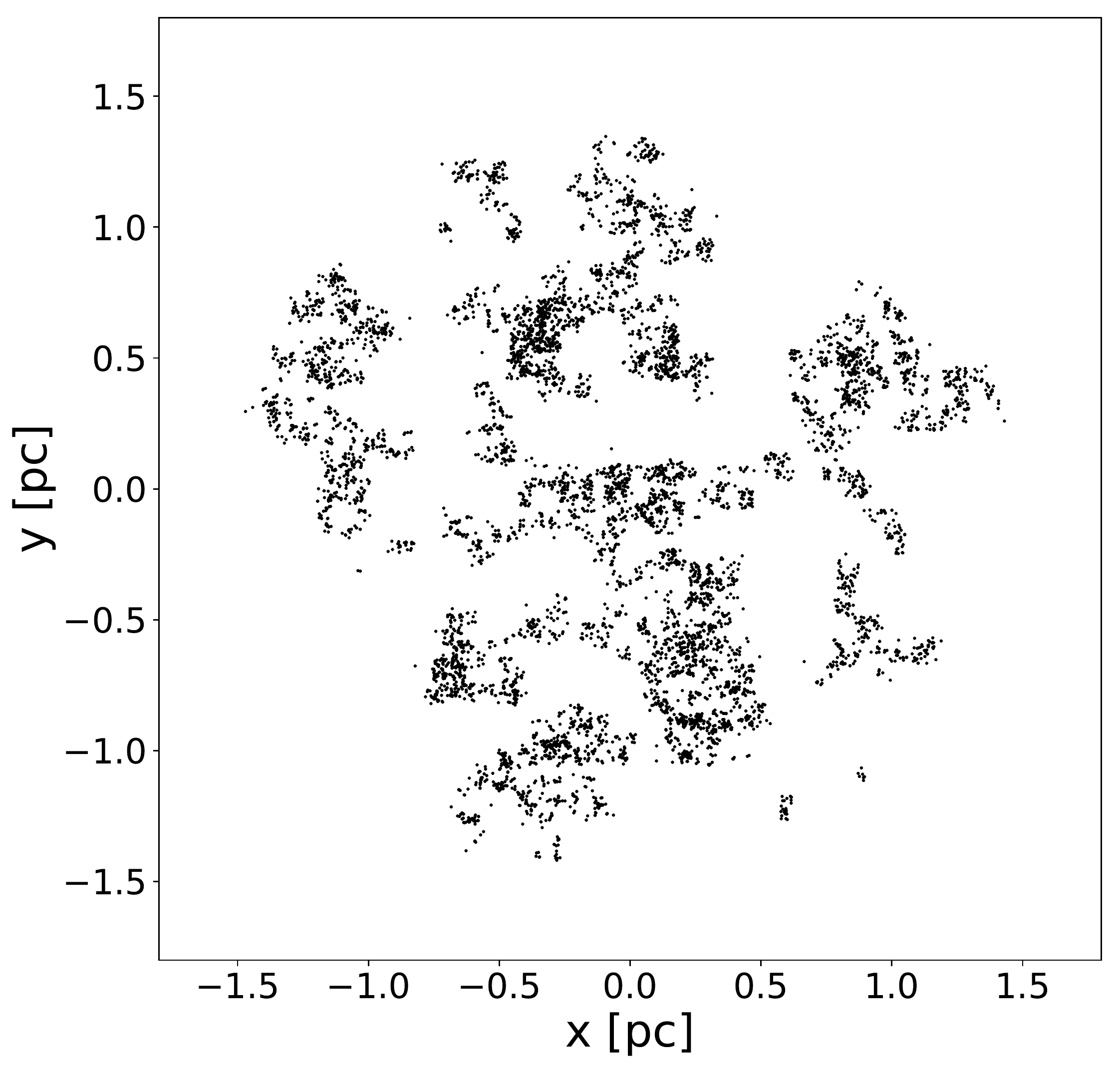,width=7.0cm} 
    \epsfig{figure=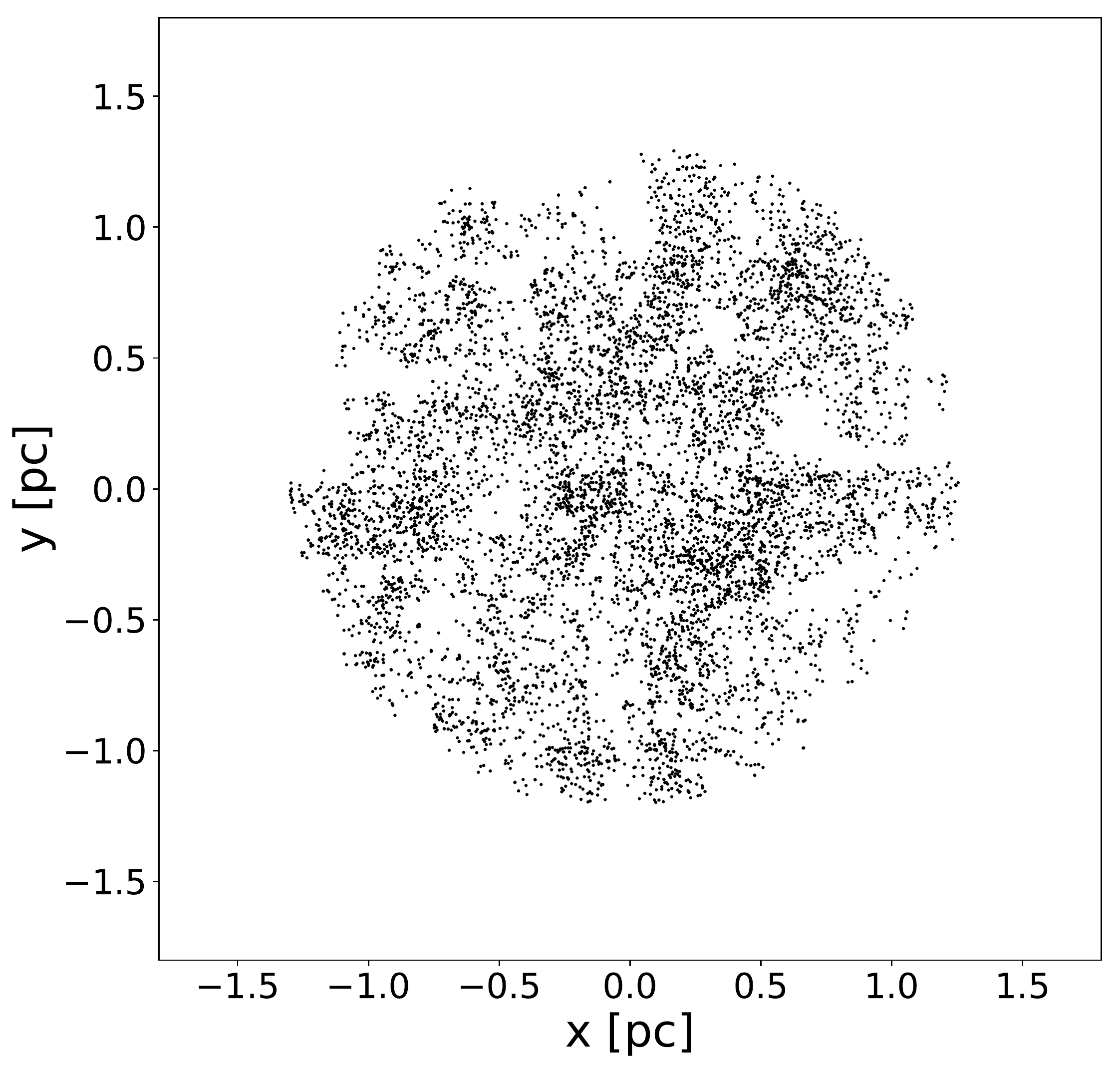,width=7.0cm} 
  \caption{  \label{fig:fig1}
Star cluster (SC) models generated with the \textsc{McLuster} code. Each point represents a star. Both SCs have a total mass of 5000 \msun{}. \textbf{Top}: model with fractal dimension $D=1.6$. \textbf{Bottom}: model with fractal dimension $D=2.3$.
}}
\end{figure}

We generate the initial conditions with  \textsc{McLuster}  \citep{kuepper2011}, which allows to produce models with different degrees of fractal substructures \citep{goodwin2004}. The level of fractality is decided by the parameter $D$ (where $D=3$ means homogeneous distribution of stars). In this work, we adopt $D=1.6$ (high-fractality runs\footnote{\cite{portegieszwart2016} suggest that the fractal dimension of young SCs should be $\sim 1.6$ by comparing observations of the Orion Trapezium Cluster with N-body simulations.}) and $D=2.3$ (low-fractality runs).  The qualitative difference between these two fractal dimensions is represented in Figure \ref{fig:fig1}.

In this work we have simulated $4\times 10^3$ fractal young SCs: half of them with $D=1.6$ (HF sample) and the remaining half with $D=2.3$ (LF sample). The total mass $M_{\rm SC}$ of each SC (ranging from $1000$ \msun{} to $30000$ \msun{})  is drawn from a distribution $dN/dM_{\rm SC}\propto M_{\rm SC}^{-2}$, as the embedded SC mass function described in \cite{lada2003}. Thus, our simulated SCs represent a synthetic young SC population of Milky Way-like galaxies.

We choose the initial SC half mass radius $r_h$ according to the Marks \& Kroupa relation \citep{marks2012}, which relates the total mass of the SC $M_{\mathrm{SC}}$ with its initial half mass radius $r_h$:
\begin{equation}
r_h=0.10^{+0.07}_{-0.04}\,{}{\rm pc}\,{} \left( \frac{M_{\mathrm{SC}}}{M_{\odot}}\right)^{0.13\pm 0.04}.
\end{equation}
The SCs are initialised in virial equilibrium.

All simulated SCs have stellar metallicity $Z=0.002$, i.e. $1/10$ of the solar metallicity (if we assume solar metallicity $Z_{\odot{}}=0.02$). The stars in the simulated SCs follow a \cite{kroupa2001} initial mass function, with minimum mass 0.1 \msun{}  and maximum mass 150 \msun{}. 

We assume an initial binary fraction $f_{\mathrm{bin}}=0.4$. The orbital periods, eccentricities and mass ratios of binaries are drawn from \cite{sana2012}. In particular, we obtain the mass of the secondary $m_{\mathrm{2}}$ as follows:
\begin{equation}
	\mathfrak{F}(q)~ \propto ~q^{-0.1} \qquad ~~~\mathrm{with}~~~q = \frac{m_2}{m_1}~ \in [0.1-1]~,
\end{equation}
the orbital period $P$ from
\begin{equation}
	\mathfrak{F}(\mathscr{P}) ~\propto~ (\mathscr{P})^{-0.55} ~~\mathrm{with}~ \mathscr{P} = \mathrm{log_{10}}(P/\mathrm{day}) \in [0.15-5.5]
\end{equation} 
and  the eccentricity $e$ from
\begin{equation}
	\mathfrak{F}(e) ~\propto ~e^{-0.42} \qquad ~~\mathrm{with}~~~ 0\leq e < 1.~
\end{equation} 
 In a forthcoming paper we will also investigate the effects of different metallicities, virial ratios and initial binary fractions.

The force integration includes a solar neighbourhood-like static external tidal field. In particular, the simulated SCs are assumed to be on a circular orbit around the center of the Milky Way with a semi-major axis of $8\,\mathrm{kpc}$ \citep{wang2016}.
Each SC is evolved until its dissolution or for a maximum time $t=100\,\mathrm{Myr}$.
The choice of terminating the simulations at $t=100\,\mathrm{Myr}$ is motivated by the fact that our tidal field model tends to overestimate the lifetime of SCs, because we do not account for massive perturbers (e.g. molecular clouds), which accelerate the SC disruption \citep{gieles2006}.

We compare the results of the SC simulations with isolated binary simulations performed with the standalone version of \textsc{MOBSE}. In particular, we simulate $10^7$ isolated binaries (IBs) with the same initial conditions as SC binaries, i.e. metallicity $Z=0.002$, primary mass drawn from a Kroupa \citep{kroupa2001} mass function between 0.1 and 150 M$_\odot$, secondary mass, eccentricity and orbital periods derived from \cite{sana2012}.

A summary of the initial conditions of the performed simulations is reported in Table \ref{tab:table1}. In the following, we consider four simulation sets: i) HF: simulated SCs with high level of fractality ($D=1.6$); ii) LF: simulated SCs with low level of fractality ($D=2.3$); iii) SC: all simulated SCs (considering HF and LF simulations together); iv) IB: isolated binary simulations run with the standalone version of \textsc{MOBSE}.

\begin{table*}
\begin{center}
\caption{\label{tab:table1} Initial conditions.} \leavevmode
\begin{tabular}[!h]{cccccccccc}
\hline
Set & Run number & $M$ [M$_\odot{}$] & $r_h$ [pc] & $Z$ & $f_{\mathrm{bin}}$ & $D$ & IMF & $m_{\mathrm{min}}$ [M$_\odot{}$] & $m_{\mathrm{max}}$ [M$_\odot{}$]
\\
\hline
HF & $2\times 10^3$   & $10^3-3\times 10^4$ & $0.1\times \left( M_{\mathrm{SC}}/M_{\odot}\right)^{0.13}$ & 0.002 & 0.4 & 1.6 & Kroupa (2002) & 0.1 & 150\\
LF & $2\times 10^3$   & $10^3-3\times 10^4$ & $0.1\times \left( M_{\mathrm{SC}}/M_{\odot}\right)^{0.13}$ & 0.002 & 0.4 & 2.3 & Kroupa (2002) & 0.1 & 150\\
SC & $4\times{}10^3$  & $10^3-3\times 10^4$ & $0.1\times \left( M_{\mathrm{SC}}/M_{\odot}\right)^{0.13}$ & 0.002 & 0.4 & 1.6, 2.3 & Kroupa (2002) & 0.1 & 150\\
IB & $10^7$        & --                 &  --                                                 & 0.002 & 1.0 & --  & Kroupa (2002) & 5 & 150\\
\hline
\end{tabular}
\end{center}
\begin{flushleft}
\footnotesize{Column~1: Name of the simulation set. HF: high fractality ($D=1.6$) N-body simulations; LF: low fractality ($D=2.3$) N-body simulations; SC: all N-body simulations considered together (i.e. set HF + set LF); IB: isolated binaries (population synthesis simulations run with {\sc MOBSE}, without dynamics). Column~2: Number of runs; column~3: total mass of SCs ($M$); column~4: half-mass radius ($r_{\mathrm{h}}$); column~5: metallicity ($Z$); column~6: initial binary fraction ($f_{\mathrm{bin}}$); column~7: fractal dimension ($D$); column~8: initial mass function (IMF); column~9: minimum mass of stars ($m_{\mathrm{min}}$); column~10: maximum mass of stars ($m_{\mathrm{max}}$).}
\end{flushleft}
\end{table*}

\section{Results}


\subsection{Statistics of dynamically formed double BHs (BBHs)}
First, we estimate how many double black hole binaries (BBHs) form via dynamical channels in our simulations. We define {\it exchanged binaries} as binaries formed via dynamical exchanges, whereas those binaries which were generated in the initial conditions  are dubbed as {\it original binaries}\footnote{Usually, dynamicists use the adjective {\it primordial} to refer to binaries which are already present in the initial conditions, but we use a different adjective to avoid confusion with primordial BHs (i.e. BHs which originate from gravitational instabilities in the early Universe).}. At the end of the simulations ($t=100$ Myr),  $\sim{}85-89$ \%{} of all surviving BBHs are exchanged binaries. Thus, dynamical exchanges give an important contribution to the population of BBHs. We find no significant differences between runs with high fractal number (HF) and with low fractal number (LF). Table~\ref{tab:table2} shows the statistics for the different simulation sets.

We check how many of these BBHs are expected to merge by gravitational wave decay within a Hubble time ($t_{\rm H}=14$ Gyr). No binaries merge during the dynamical simulations. Thus, to calculate the merging time, we use the following equation \citep{peters1964}
\begin{equation}
t_{\mathrm{GW}}=\frac{5}{256}\frac{c^5\,{}a^4\,{}(1-e^2)^{7/2}}{G^3\,{}m_1\,{}m_2\,{}(m_1+m_2)},
\label{eq:tgw}
\end{equation}
where $c$ is the speed of light, $G$ is the gravitational constant, $e$ is the eccentricity of the binary, $a$ is the semi-major axis of the binary, $m_1$ and $m_2$ are the masses of the primary and of the secondary BH, respectively. For each binary, we take the values of $a$ and $e$ at the end of the simulation (during the simulation, {\sc MOBSE} integrates the change of eccentricity and semi-major axis driven by gravitational wave decay using a similar formalism, derived from \citealt{peters1964}).

The number of BBHs merging within a Hubble time (hereafter, merging BBHs) is also shown in Table~\ref{tab:table2}. More than half of merging BBHs ($\sim{}50$ \% and $\sim{}56$ \% in HF and LF simulations, respectively) are exchanged binaries, confirming the importance of dynamical exchanges for merging BBHs. Moreover, only 4 merging BBHs ($\lesssim{}2$ per cent of all merging BBHs) are still members of their parent SC at the end of the simulation, while the others have all been dynamically ejected.

These results confirm the importance of dynamics, and in particular of dynamical exchanges, for the demography of BBHs in small young SCs. Interestingly, the vast majority (if not all) of dynamically formed merging BBHs are no longer members of their parent SC when they merge, because of dynamical ejections. 

Overall, the populations of merging BBHs in HF and LF simulations show a similar trend. For this reason, in the rest of the paper, we will consider HF and LF simulations as a single simulation set (SC simulations), unless otherwise specified. We will discuss the differences between HF and LF simulations in Section~\ref{sec:fractality}.

\subsection{BH mass distribution}
The next step is to understand whether BBHs formed in young SCs have distinctive features with respect to isolated binaries. In this section, we consider the mass of BHs.
\subsubsection{BH mass versus ZAMS mass}
\begin{figure*}
  \center{
    \epsfig{figure=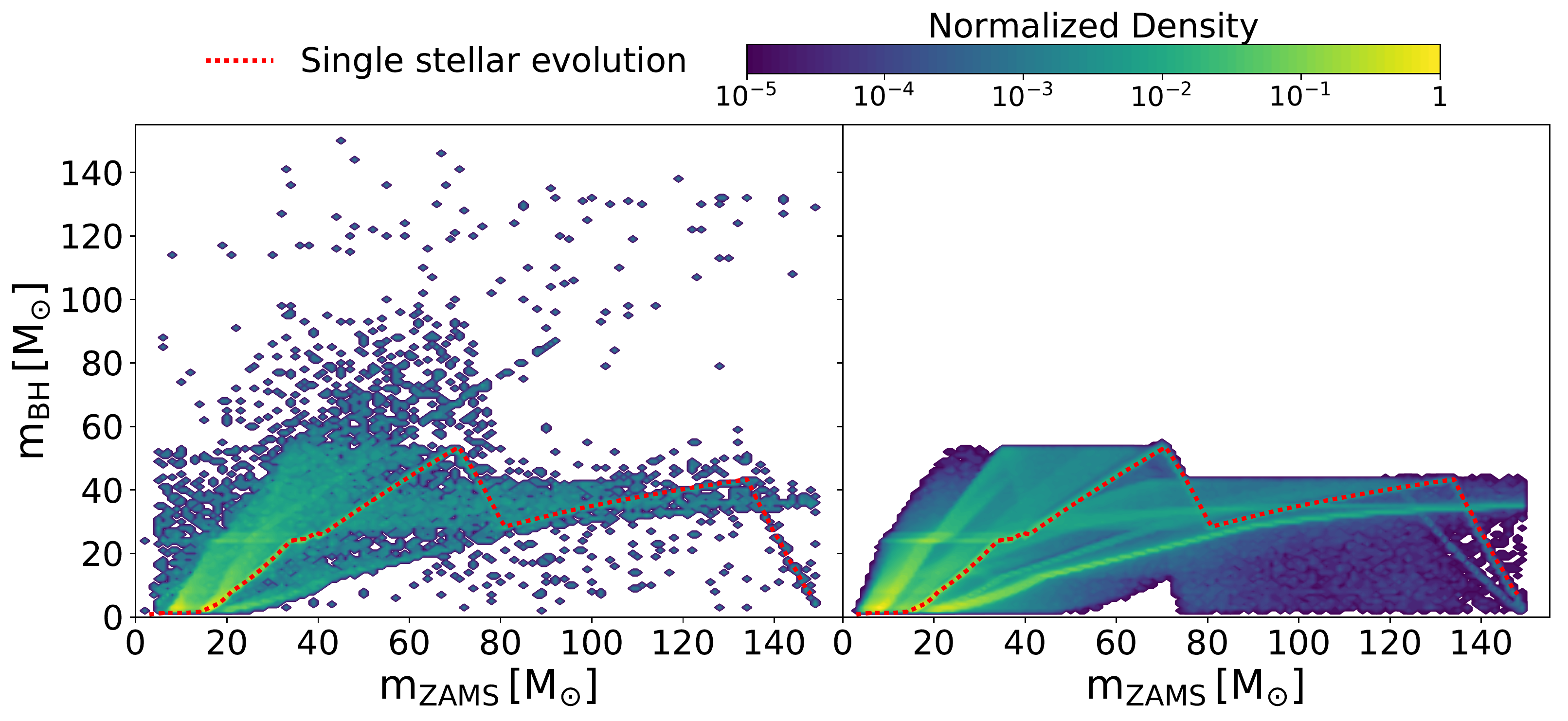,width=15.0cm} 
  \caption{BH mass ($\mathrm{m_{BH}}$) as a function of the zero-age main sequence (ZAMS) mass of the progenitor stars ($\mathrm{m_{ZAMS}}$) in the SC simulations (left), and in IBs evolved with {\sc MOBSE} (right). The logarithmic colour bar represents the number of BHs per cell, normalized to the maximum cell-value of each plot. Each cell is a square with a side of $1.5\,$\msun{}. The red dashed line represents the mass spectrum of compact objects obtained from single stellar evolution. Dynamical interactions trigger the formation of more massive BHs. Intermediate-mass BHs (IMBHs, i.e. BHs with mass $m_{\rm BH}>150$ M$_\odot$) are not shown in this Figure.\label{fig:mobdyn}}}
\end{figure*}

Figure \ref{fig:mobdyn} shows the mass of all simulated BHs with mass $m_{\rm BH}<150$ M$_\odot$ (both single and binary BHs) as a function of the zero-age main sequence (ZAMS) mass of the progenitor star. The left-hand panel shows the results of the SC simulations (considering BBHs from models HF and LF together, because the level of fractality does not significantly affect the masses, see Table~\ref{tab:table3}), while the right-hand panel shows the comparison sample of isolated binaries (IB). The BH mass we obtain from the evolution of single stars is shown as a red dotted line in both plots.

\begin{figure}
  \center{
    \epsfig{figure=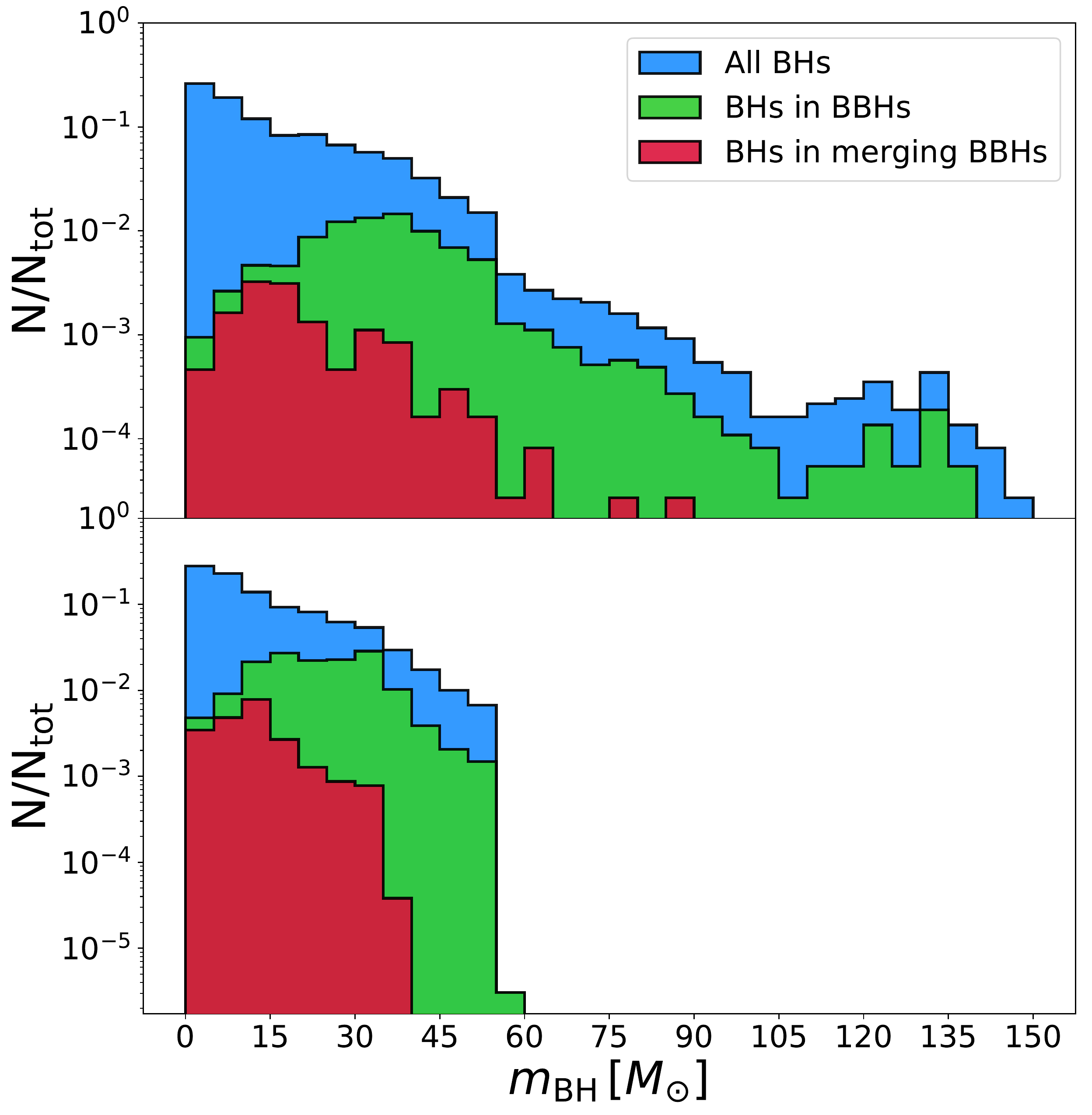,width=8.0cm} 
  \caption{Top (Bottom) panel: distribution of BH masses in the SC (IB) simulations. Blue: all BHs with mass $m_{\rm BH}<150$ M$_\odot$ (IMBHs are excluded); green: BHs which reside in BBH systems at the end of the simulations; red: BHs in merging BBH systems. \label{fig:masshist}}
}
\end{figure}

\begin{figure}
  \center{
    \epsfig{figure=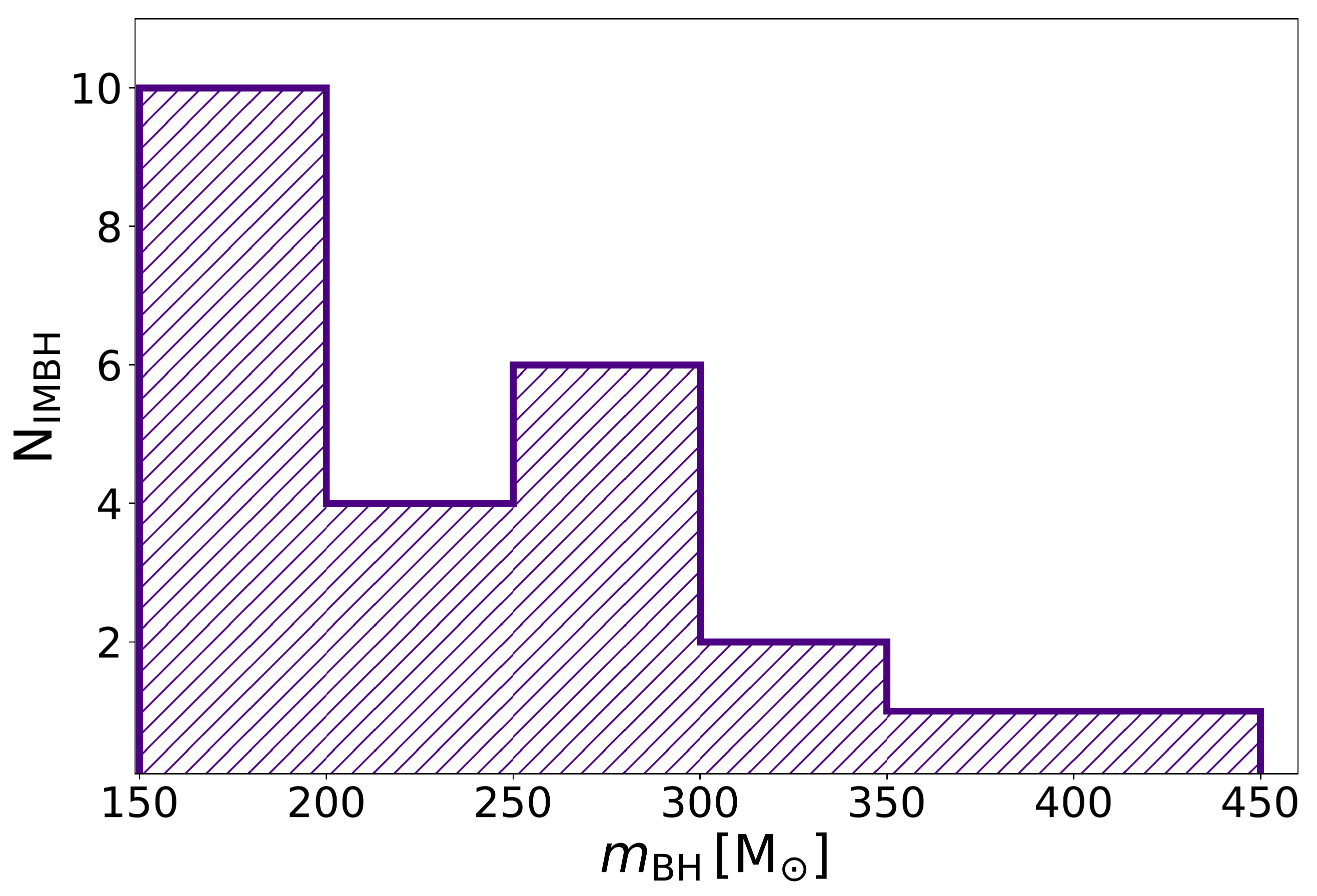,width=8.0cm} 
  \caption{Mass distribution of the 24 intermediate-mass BHs (IMBHs) formed by runway collisions in our SC simulations. \label{fig:IMBH}}
}
\end{figure}

Dynamics does not affect significantly the mass of the majority of BHs: 
 the most densely populated regions of the plot are the same in both panels. Binary evolution processes (especially mass transfer and common envelope) have a much stronger impact than dynamics on the mass of most BHs. For example, the large number of BHs with mass higher than expected from single star evolution (red dotted line) for progenitors with ZAMS mass $M_{\rm ZAMS}<60$ M$_\odot$ is an effect of mass accretion, while most of the BHs with mass lower than expected from single star evolution originate from donor stars whose envelope was peeled off (see \citealt{giacobbo2018b} and \citealt{spera2018} for more details on the effects of binary evolution on BH masses).

 However, dynamics crucially affects the maximum mass of BHs (see also Fig.~\ref{fig:masshist}). In the N-body simulations, BHs with mass up to $\sim{}440$ $M_\odot$  are allowed to form, while the maximum mass of BHs formed from isolated binary evolution is $\sim{}60$ M$_\odot$. In figures~\ref{fig:mobdyn} and \ref{fig:masshist}, we show only BHs\footnote{Second-generation BHs (\citealt{gerosa2017}, i.e. BHs formed from the merger of two BHs) are not shown in Figures~\ref{fig:mobdyn} and \ref{fig:masshist}.} with mass $<150$ M$_\odot$ for clarity. BHs with mass $>150$ M$_\odot$ are very rare  and including them would make these figures more difficult to read. Figure~\ref{fig:IMBH} shows the BHs with mass $m_{\rm BH}\ge{}150$ M$_\odot$.  These are intermediate-mass black holes (IMBHs) and form by runaway collisions of stars in the early evolution of the SC (see e.g. \citealt{portegieszwart2002,portegieszwart2004,giersz2015,mapelli2016}). In our simulations, we find 24 IMBHs, which represent only the 0.065 \% of all simulated BHs.


 The percentage of BHs with mass $>70$ M$_\odot$ in the N-body simulations is only $\sim{}1$ \% of all BHs ($0.92$~\%{} and $\sim{}0.96$ \%{} in LF and HF simulations, respectively). Thus, they are very rare, but their large mass is a clear signature of dynamical origin. These massive BHs form because of multiple stellar mergers, which can happen only in SCs. In fact, if the two members of an isolated binary merge together, the probability that their merger product acquires a new companion is negligible; whereas the merger product of a binary in a SC has a good chance to acquire a new companion dynamically (especially if it is particularly massive) and might merge also with new companion \citep{mapelli2016}. Of course, these multiple mergers are mostly mergers of stars, because mergers of BBHs are much rarer events (see next section). In particular, the most massive BHs in our simulations form from the merger of at least four stars.

Because the most massive BHs  originate from multiple stellar mergers (or even runaway collisions), the maximum BH mass in the N-body simulations essentially does not depend on the ZAMS mass of the progenitor (in the case of mergers, Figure~\ref{fig:mobdyn} shows the ZAMS mass of the most massive among the stellar progenitors).

We expect PISNe and PPISNe to prevent the formation of BHs with mass $\sim{}60-120$ M$_\odot$ from single stellar evolution \citep{belczynski2016pair,spera2017,woosley2017}. On the other hand, \cite{spera2018} have already shown that binary evolution might produce few BHs in the mass gap. For example, the merger between an evolved star (with an already well developed Helium or Carbon-Oxygen core) and a main sequence star might produce a very massive star with a large Hydrogen envelope but with a Helium core smaller than required to enter the pair-instability range. Such star might end its life directly collapsing to a BH with mass in the pair instability gap. In a SC, such massive BHs have additional chances to form (see e.g. \citealt{mapelli2016}), because multiple mergers between stars are likely.

Moreover, if a BH with mass $\sim{}60-120$ M$_\odot$ forms from the merger of an isolated binary, it will remain a single BH. In contrast, if such BH forms in a SC, it might acquire another companion by dynamical exchange and it might merge by gravitational wave emission.

\begin{table*}
\begin{center}
\caption{\label{tab:table2} Number of SC BBHs.} \leavevmode
\begin{tabular}[!h]{ccccc}
\hline
Set & Exchanged BBHs & Original BBHs & Merging Exchanged BBHs & Merging Original BBHs\\
\\
\hline
HF & 710 & 124 & 60 & 59\\
LF & 786 & 98 & 62 & 48\\
\hline
\end{tabular}
\end{center}
\begin{flushleft}
  \footnotesize{Number of BBHs in HF and LF simulations.    Column~1: Name of the simulation set; column~2: number of BBHs that formed from dynamical exchanges and that are still bound at the end of the simulations ($t=100$ Myr); column~3: number of  BBHs that formed from original binaries and that are still bound at the end of the simulations ($t=100$ Myr); column~4: number of BBHs  that formed from dynamical exchanges and that merge within a Hubble time (hereafter merging exchanged BBHs); column~5: number of BBHs that formed from original binaries and that merge within a Hubble time (hereafter merging original BBHs).}
\end{flushleft}
\end{table*}

\subsubsection{Mass of single versus binary BHs}

The top panel of Figure \ref{fig:masshist} compares the mass distribution of all BHs with $m_{\rm BH}<150$ M$_\odot$ formed in the N-body simulations (including merger products) with the mass of BHs which are members of BBHs by the end of the N-body simulations. Light BHs are less likely to be members of a binary than massive BHs: just $\sim{}4$ \% of BHs with mass lower than $30\,$\msun{} reside in BBHs by the end of the SC simulations, while $\sim{}29$ \% of BHs with mass larger than $30\,$\msun{} have a BH companion by the end of the SC simulations. 


The bottom panel of Figure \ref{fig:masshist} shows the same quantities for the IB simulations. The difference between top and bottom panel is apparent: no BHs with mass larger than 60 M$_\odot$ form in the IB simulations at metallicity $Z=0.002$, while SC dynamics induces the formation of BHs with mass up to $\sim{}440$ M$_\odot$ (Fig.~\ref{fig:IMBH}).


We stress that BHs with mass $\gtrsim{}60$ M$_\odot$ are inside the mass gap expected from PISNe and PPISNe \citep{belczynski2016,woosley2017,spera2017}: they can form dynamically because of multiple mergers between stars.

\subsubsection{Mass of merging BHs: exchanged, original and isolated binaries}
From Fig.~\ref{fig:masshist} it is apparent that the most massive BHs do not merge within a Hubble time, in both IBs and SCs. This effect was already discussed in previous work \citep{giacobbo2018,giacobbo2018b,spera2018} and is a consequence of the stellar radius evolution of the progenitors of such massive BHs.\footnote{Unless chemical homogeneous evolution is assumed, the radii of massive stars ($m_{\rm ZAMS}\gtrsim{}40$ M$_\odot$) reach values of several hundred to several thousand R$_\odot$. Stellar binaries with initial orbital separation larger than these radii survive early coalescence and can evolve into massive BBHs, but the latter cannot merge via gravitational wave decay because their semi-major axis is too large. In contrast, stellar binaries with smaller orbital separations either merge before becoming BHs or undergo a non-conservative mass transfer (or common envelope) process. At the end of this process, the binary might survive and evolve into a BBH, but the mass of the two BHs will be significantly smaller than expected from single stellar evolution, because of mass loss during mass transfer \citep{giacobbo2018b,spera2018}.}

However, if we compare the two red histograms in the top and bottom panel, we see that the maximum mass of  merging BHs is larger in SCs than in IBs. Thus, dynamics triggers the merger of some massive ($>40$ M$_\odot$) BHs, which cannot merge if they evolve in IBs. 

\begin{figure}
  \center{
    \epsfig{figure=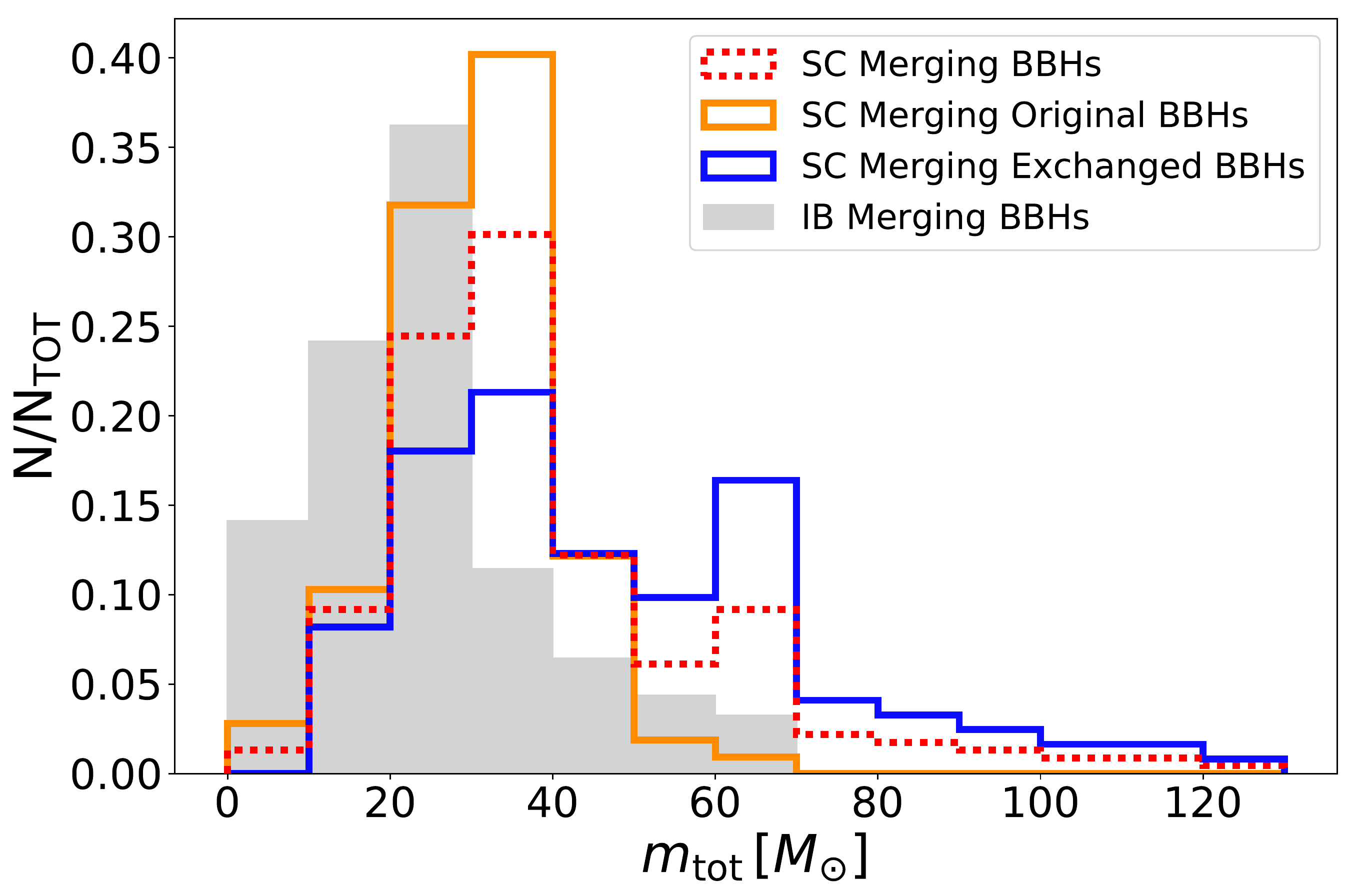,width=8.0cm} 
  \caption{Distribution of total masses ($m_{\rm tot}=m_1+m_2$) of merging BBHs. Orange solid line: original BBHs formed in SCs; blue solid line: exchanged BBHs formed in SCs; red dashed line: all merging BBHs formed in SCs (original+exchanged); gray filled histogram: BBHs formed in isolated binaries (IBs). Each distribution is normalized to its total number of elements. \label{fig:totalmasses}}
}
\end{figure}

\begin{figure}
  \center{
    \epsfig{figure=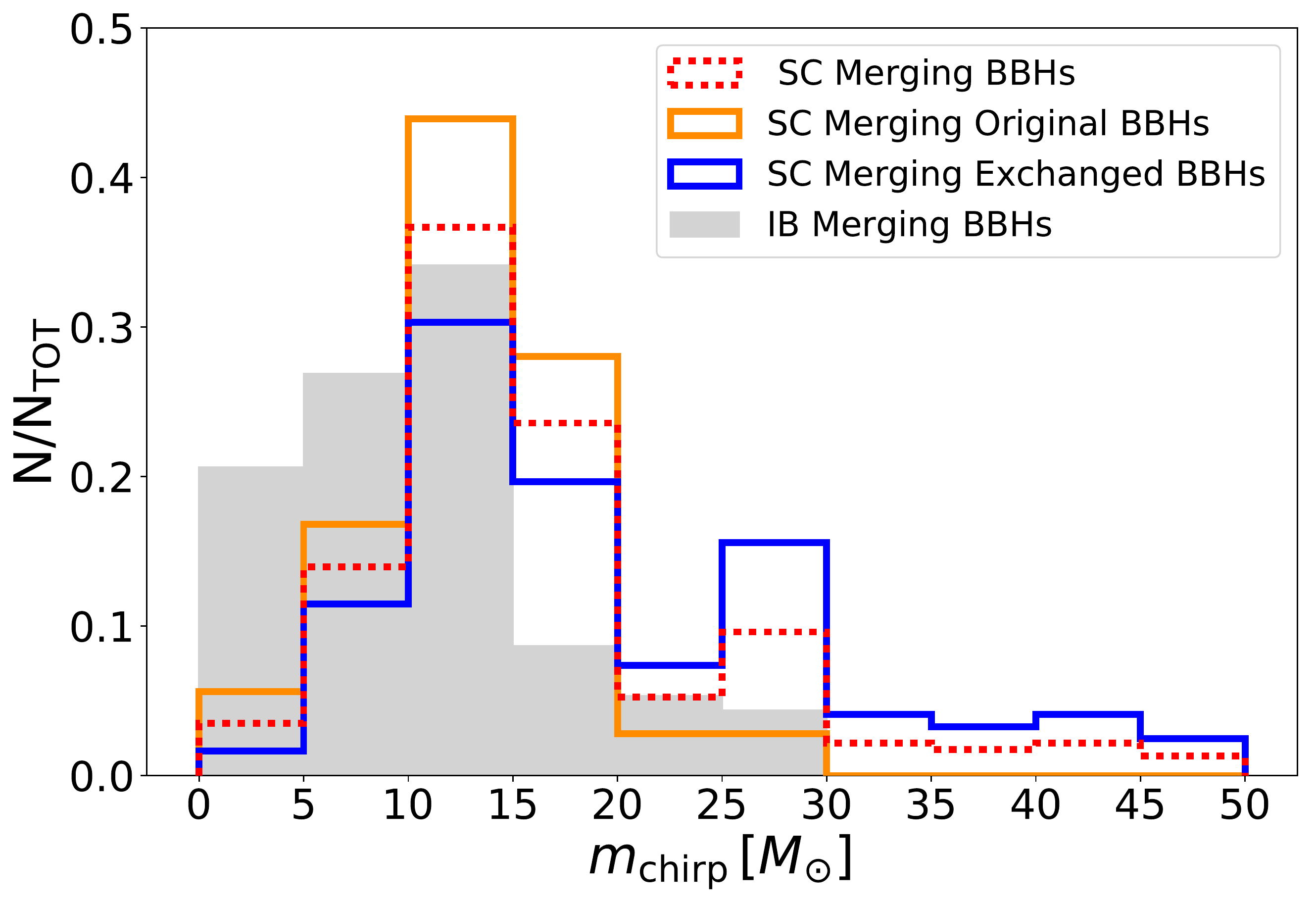,width=8.0cm} 
  \caption{Same as Figure~\ref{fig:totalmasses}, but for the distribution of chirp masses $m_{\rm chirp}=(m_1\,{}m_2)^{3/5}(m_1+m_2)^{-1/5}$ of merging BBHs. \label{fig:chirpmass}}
}
\end{figure}

\begin{figure}
  \center{
    \epsfig{figure=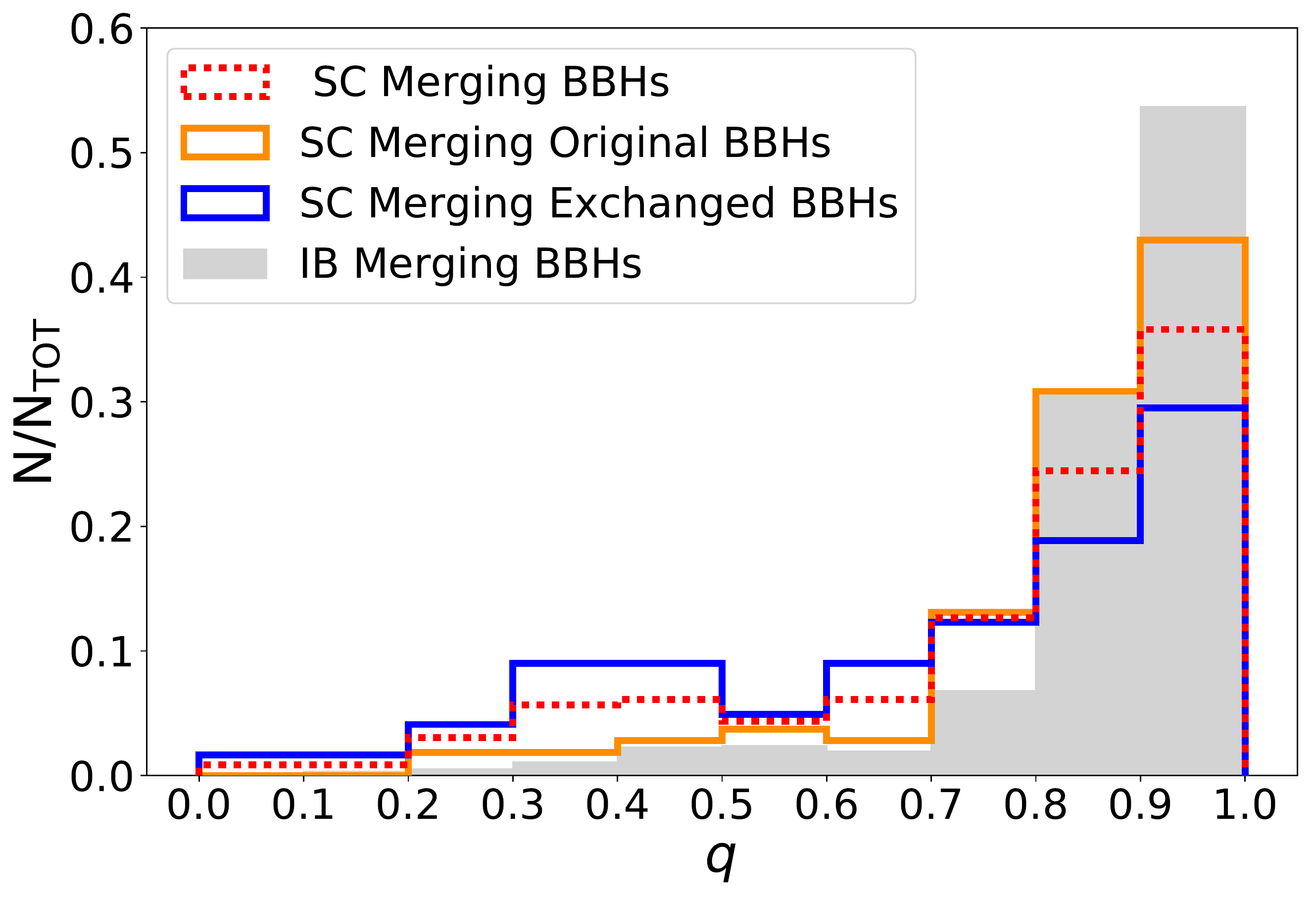,width=8.0cm} 
  \caption{Same as Figure~\ref{fig:totalmasses}, but for the distribution of mass ratios $q=m_2/m_1$ of merging BBHs. \label{fig:massratios}}
}
\end{figure}

Figure \ref{fig:totalmasses} shows the total masses ($m_{\rm tot}=m_1+m_2$) of merging exchanged and original BBHs in SCs. The former are merging BBHs which form from exchanged binaries in SC simulations, while the latter are merging BBHs which form from original binaries in SC simulations. For comparison, we also show the total masses of merging BBHs from IB simulations. The distribution of exchanged BBHs is markedly different from the other two. In particular, merging exchanged BBHs tend to be more massive than original BBHs: $m_{\rm tot}\leq{}70$ M$_\odot$ for both isolated BBHs and original BBHs,  while merging exchanged BBHs can reach $m_{\rm tot}\sim{}120$ M$_\odot$. Table~\ref{tab:table3} shows the results of the Wilcoxon U-test \citep{bauer1972,wolfe1999} and of the Kolmogorov-Smirnov (KS) test \citep{birnbaum1951,wang2003}, which confirm that the mass distribution of merging exchanged BBHs is not consistent with the other two classes of merging BBHs.

This large difference can be easily explained with the properties of dynamical exchanges: exchanges tend to favour the formation of more and more massive binaries, because these are more energetically stable (see e.g. \citealt{hills1980}).

Interestingly, the mass distribution of merging original BBHs 
is also significantly different from the distribution of merging BBHs formed from IB simulations (see Table~\ref{tab:table3}). Since both IBs and original binaries in SCs were evolved with the same population synthesis code ({\sc MOBSE}), dynamical effects are the only way to explain this difference. Even if they do not form through dynamical exchanges, also original BBHs participate in the dynamical life of a SC: they can be hardened or softened or even ionized by three-body encounters.  More massive BBHs are more likely to be hardened by three-body encounters, while light BBHs are more likely to be softened or ionized. This explains why merging original BBHs tend to be more massive than  merging BBHs in IBs. Thus, dynamics affects almost the entire population of merging BBHs in SCs: not only exchanged BBHs, but also original BBHs.

Similar considerations apply also to the distribution of chirp masses $m_{\rm chirp}=(m_1\,{}m_2)^{3/5}(m_1+m_2)^{-1/5}$ (Figure~\ref{fig:chirpmass}): we find no merging original BBHs with $m_{\rm chirp}>30$ M$_\odot$, while merging exchanged BBHs reach $m_{\rm chirp}\sim{}50$ M$_\odot$.

Figure~\ref{fig:massratios} shows the mass ratios of merging BBHs (defined as $q=m_2/m_1$). All distributions peak at $q\sim{}1$. However, small mass ratios are significantly more likely in merging exchanged  BBHs than in both merging original and isolated BBHs. 
Table~\ref{tab:table3} confirms that the distribution of $q$ of the merging exchanged BBHs is significantly different from both merging original BBHs and isolated binaries. This can be easily explained with the different formation channels. If two BHs form from the same close stellar binary, mass transfer and common envelope episodes tend to ``redistribute'' the mass inside the system, leading to the formation of two BHs with similar mass. In contrast, if two BHs enter the same binary by exchange, after they formed, their mass cannot change anymore and even extreme mass ratios $q\sim{}0.1$ are possible. 

\begin{table*}
\begin{center}
\caption{\label{tab:table3} Results of the U-Test and KS-Test to compare two BBH samples.} \leavevmode
\begin{tabular}[!h]{lllllll}
\hline
BBH sample 1 & BBH sample 2 & Distribution & U-Test & KS-Test & Median 1 & Median 2\\
\\
\hline

HF -- Merging BBHs & LF -- Merging BBHs & $M_{\rm tot}$  &  0.57 & 0.20 & 32.8 & 35.4 \\
HF -- Merging BBHs & LF -- Merging BBHs & $M_{\rm chirp}$  & 0.51 & 0.50 & 14.0 & 14.6  \\
HF -- Merging BBHs & LF -- Merging BBHs & $q$  & 0.03 & 0.05 & 0.88 & 0.81 \\

HF -- Merging BBHs & LF -- Merging BBHs & $t_{\rm exch}$  & 0.04 & 0.11 & 1.9 & 2.8  \\
\hline

SC -- Merging BBHs & IB -- Merging BBHs & $M_{\rm tot}$  &  0 & 0 & 34.7 & 24.0  \\
SC -- Merging BBHs & IB -- Merging BBHs & $M_{\rm chirp}$ & 0 & 0 & 14.2 & 10.4 \\
SC -- Merging BBHs & IB -- Merging BBHs & $q$ & $10^{-7}$ & $10^{-7}$ & 0.84 & 0.89 \\

\hline

SC -- Merging Exchanged BBHs & SC -- Merging Original BBHs  & $M_{\rm tot}$ & $10^{-8}$ & $10^{-8}$ & 41.5 & 30.2 \\
SC -- Merging Exchanged BBHs & IB -- Merging BBHs  & $M_{\rm tot}$ & 0 & 0 &  41.5 & 24 \\
SC -- Merging Original BBHs  & IB -- Merging BBHs  & $M_{\rm tot}$ & $10^{-9}$ & $10^{-9}$ & 30.2 & 24 \vspace{0.2cm}\\

SC -- Merging Exchanged BBHs & SC -- Merging Original BBHs  & $M_{\rm chirp}$ & $10^{-5}$ & $10^{-7}$ & 16.9 & 13.2 \\
SC -- Merging Exchanged BBHs & IB -- Merging BBHs & $M_{\rm chirp}$ & 0 & 0 & 16.9 & 10.4 \\
SC -- Merging Original BBHs  & IB -- Merging BBHs   & $M_{\rm chirp}$ & $10^{-9}$ & $10^{-9}$ & 13.2 & 10.4 \vspace{0.2cm}\\

SC -- Merging Exchanged BBHs & SC -- Merging Original BBHs  & $q$ & $10^{-4}$ & $10^{-4}$ & 0.78 & 0.89 \\
SC -- Merging Exchanged BBHs & IB -- Merging BBHs & $q$ & $10^{-9}$ & $10^{-9}$ & 0.78 & 0.89 \\
SC -- Merging Original BBHs  & IB -- Merging BBHs   & $q$ & 0.27 & 0.14 &  0.89 & 0.89 \\

\hline
\end{tabular}
\end{center}
\begin{flushleft}
  \footnotesize{In this Table we apply the U- and KS- tests to compare different samples of merging BBHs (i.e. BBHs merging within a Hubble time). Columns~1 and 2: the two BBH samples to which we apply the U- and KS- test. Each sample comes from one of the simulation sets (HF, LF, SC and IB, see Table~\ref{tab:table1}). 
    For the SC sample we also distinguish between ``Exchanged BBHs'' and ``Original BBHs'' (see Section 3.1 for the definition).
    Column~3: distribution to which we apply the U- and KS- tests. We consider total BBH masses ($M_{\rm tot}$), chirp masses ($M_{\rm chirp}$), mass ratios ($q$) and time of the exchange ($t_{\rm exch}$, for exchanged binaries only). 
    Columns~4 and 5: probability that the two samples are drawn from the same distribution according to the U-Test and to the Kolmogorov-Smirnov (KS) Test, respectively. Values smaller than $10^{-20}$ are indicated with 0. Columns~6 and 7: median values of the distributions of BBH sample 1 and 2, respectively.}
\end{flushleft}
\end{table*}

\begin{figure}
  \center{
    \epsfig{figure=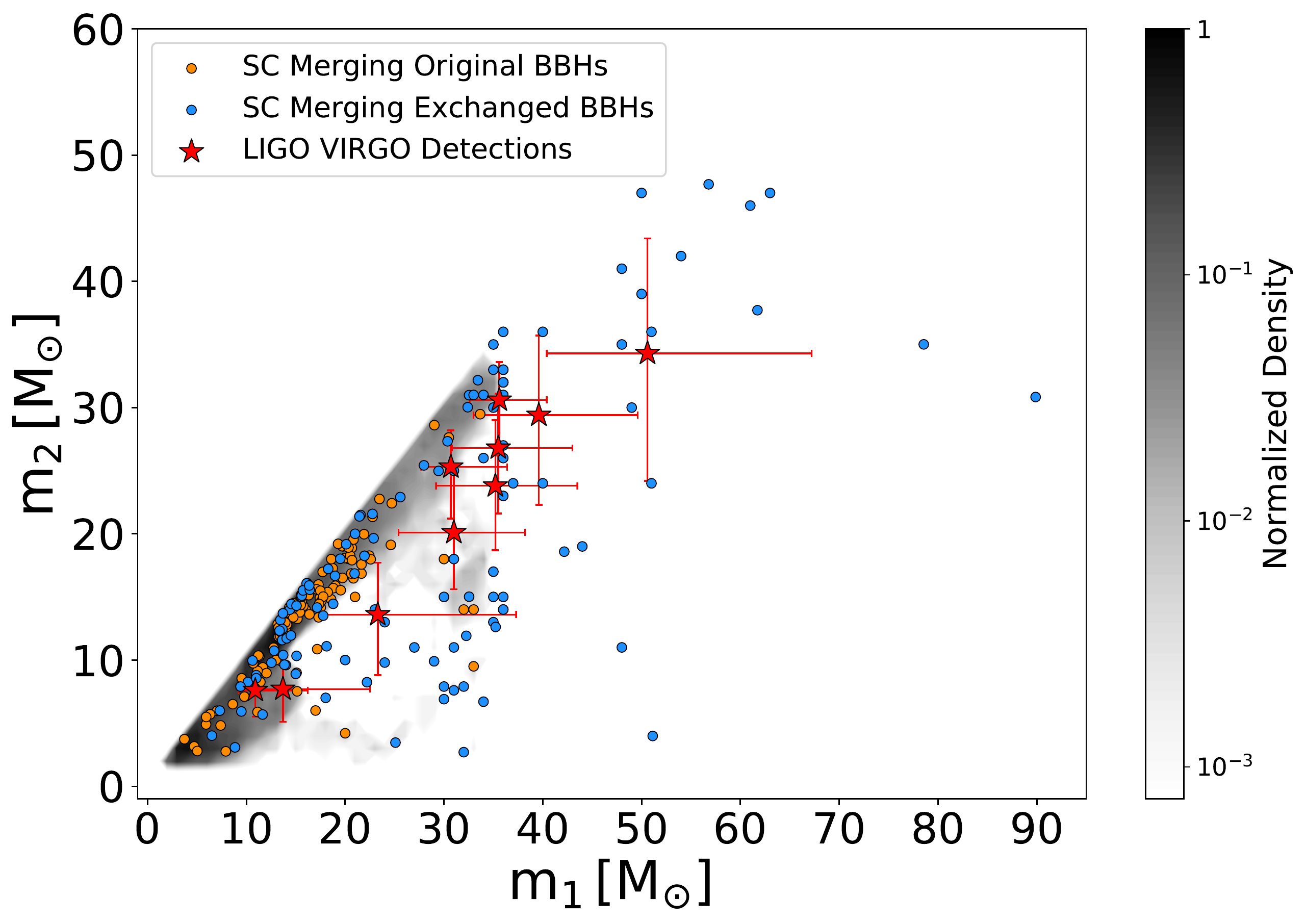,width=9.0cm} 
  \caption{ \label{fig:LIGOVirgo} Mass of the primary BH ($m_1$) versus mass of the secondary BH ($m_2$) of merging BBHs. Filled contours (with gray colour map): IBs; orange circles: SC merging original BBHs; blue circles: SC merging exchanged BBHs; red stars with error bars: LIGO-Virgo detection of BH mergers -- GW150914 \citep{abbottGW150914}, GW151012 \citep{abbottO1}, GW151226 \citep{abbottGW151226}, GW170104 \citep{abbottGW170104}, GW170608 \citep{abbottGW170608}, GW170729 \citep{abbottO2}, GW170809 \citep{abbottO2}, GW170814 \citep{abbottGW170814}, GW170818 \citep{abbottO2} and GW170823 \citep{abbottO2}. Error bars indicate 90\% credible levels.}
}
\end{figure}

Finally, Figure~\ref{fig:LIGOVirgo} shows the mass of the primary BH versus the mass of the secondary BH of merging systems. The first ten GW events associated with BBHs are also shown. It must be kept in mind that we are not weighting the simulated distributions for the probability of observing them with LIGO-Virgo (e.g. more massive BHs can be observed to a farther distance than light BHs, \citealt{abbottastrophysics}). From this Figure, it is apparent that the LIGO-Virgo BBHs lie in a region of the $m_1-m_2$ plane which is well populated by both IB and SC merging BBHs at metallicity $Z=0.002$. Merging exchanged BHs are clearly different from the other two populations, both in terms of masses and in terms of mass ratios. Interestingly, the most massive event GW170729 \citep{abbottO2} lies in a region that is populated only by merging exchanged BBHs.

\begin{table}
\begin{center}
\caption{\label{tab:table4} Heavy merging BBHs.} \leavevmode
\begin{tabular}[!h]{ccc}
\hline
$m_1$ (M$_\odot$) & $m_2$ (M$_\odot$) & $q$\\
\hline
90 & 31 & 0.34\\
79 & 35 & 0.44\\
63 & 47 & 0.75\\
62 & 38 & 0.53\\
61 & 46 & 0.75\\
57 & 48 & 0.84\\
54 & 42 & 0.78\\
51 & 36 & 0.71\\
51 & 24 & 0.47 \\
50 & 47 & 0.94\\
50 & 39 & 0.78\\
48 & 41 & 0.85\\
48 & 35 & 0.73\\
49 & 30 & 0.61\\
40 & 36 & 0.78\\
36 & 36 & 1.00\\
35 & 35 & 1.00\\
\hline
\end{tabular}
\end{center}
\begin{flushleft}
  \footnotesize{Properties of merging BBHs with total mass $\ge{}70$ M$_\odot$. Column~1: primary mass ($m_1$); column~2: secondary mass ($m_2$); column~3: mass ratio ($q$). All these binaries are exchanged BBHs.}
\end{flushleft}
\end{table}

Table~\ref{tab:table4} summarizes the masses of the most massive merging BBHs in our simulations (with $m_{\rm tot}\ge{}70$ M$_\odot$). All of them are SC exchanged BBHs. The mass of the primary spans from 35 to 90 M$_\odot$, with five BHs more massive than 60 M$_\odot$ (these lie in the mass gap produced by PPISNe and PISNe). The mass of the secondary ranges from 24 to 48 M$_\odot$. These massive merging BBHs have median mass ratio $q=0.75$, significantly lower than 1. The most massive binary ($m_1=90$ M$_\odot$, $m_2=31$ M$_\odot$) has also the lowest mass ratio $q=0.34$.




\subsection{Orbital eccentricity of merging BBHs}
Orbital eccentricity is another feature which can be probed with current and future gravitational-wave detectors \citep{nishizawa2016,nishizawa2017,rodriguez2018}. Dynamical exchanges tend to produce more eccentric binaries, although gravitational wave emission circularizes them very efficiently. Thus, even binaries with extreme eccentricity might achieve almost null eccentricity when they reach the LIGO-Virgo band ($>10$ Hz). However, a significant fraction of exchanged BBHs formed in globular clusters (e.g. \citealt{rodriguez2018}) or nuclear SCs (e.g. \citealt{antonini2016}) is predicted to have eccentricity $e\gtrsim{}10^{-2}$ when entering the LISA band ($\sim{}10^{-4}-1$ Hz). 

\begin{figure*}
  \center{
    \epsfig{figure=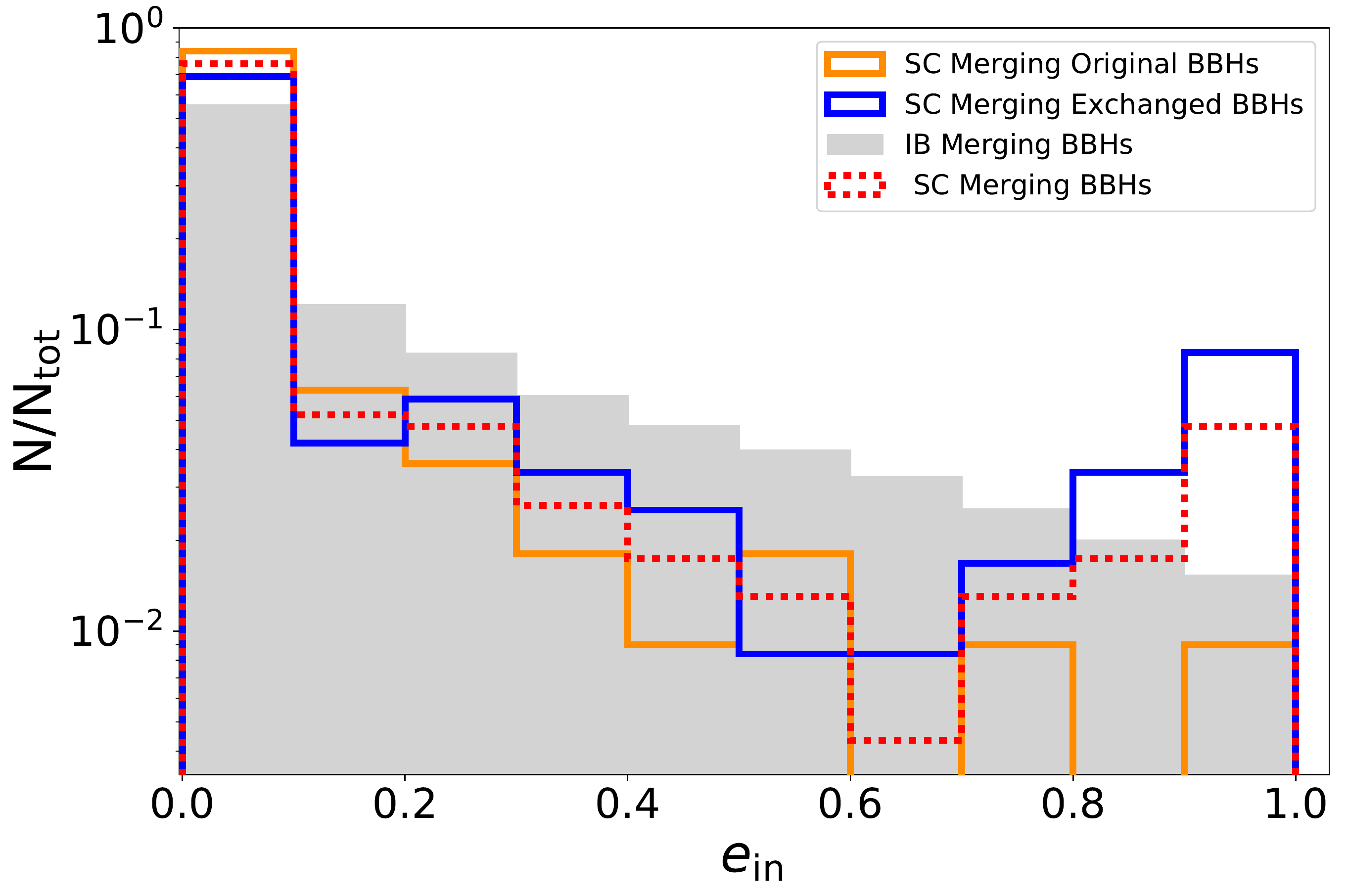,width=8.0cm} 
    \epsfig{figure=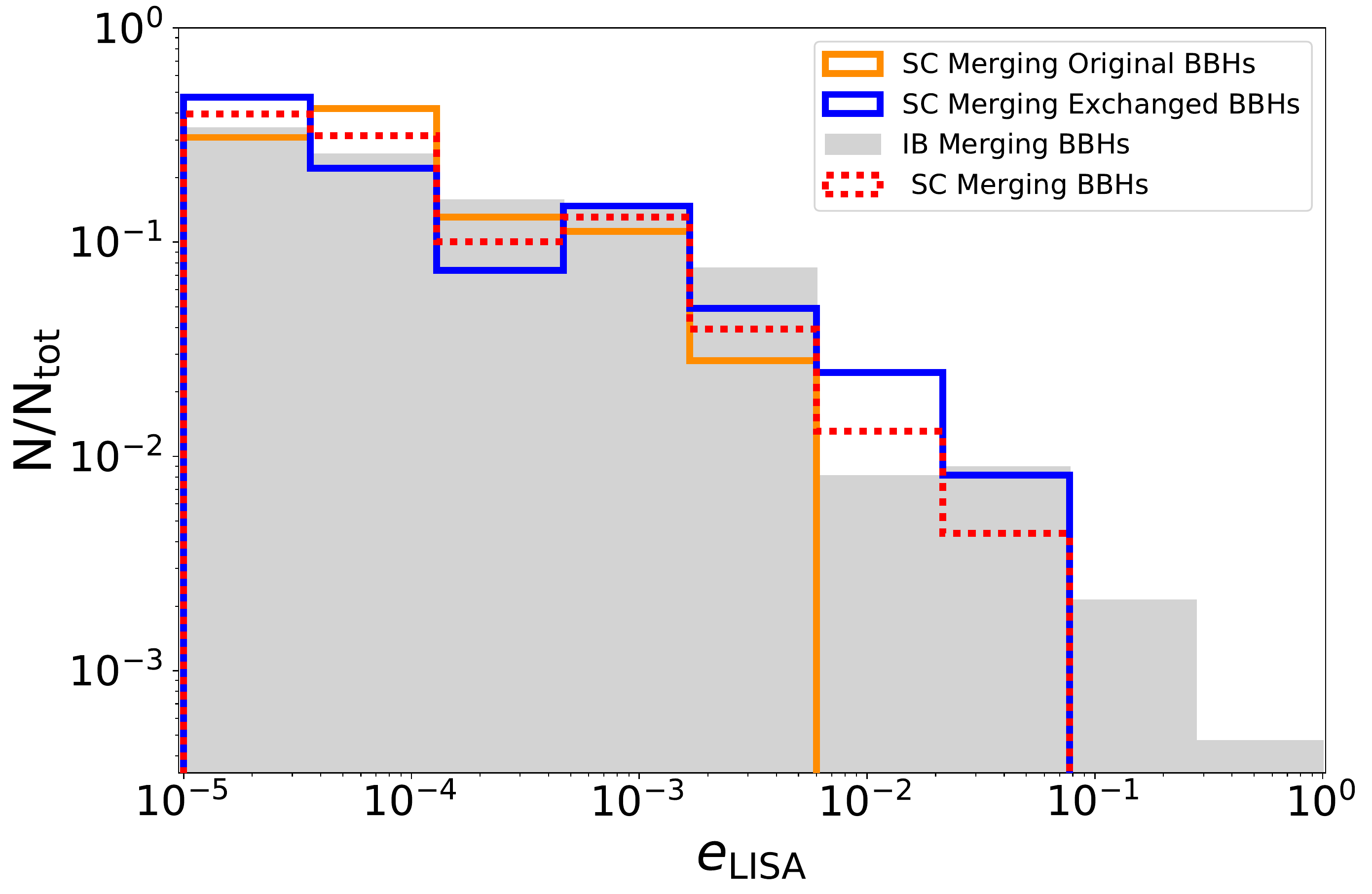,width=8.0cm} 

    \caption{Distribution of eccentricities of merging BBHs in SCs and in IBs. All merging BBHs in SCs are shown by the dashed red line. BBHs in IBs are shown by the filled gray histogram. SC merging original BBHs and SC merging exchanged BBHs are shown with an orange and blue line, respectively.  The eccentricity $e_{\rm in}$ (left) is calculated when the binary becomes a BBH, while $e_{\rm LISA}$ (right) is calculated when the orbital frequency is $f_{\rm orb}=10^{-2}$ Hz (where the sensitivity of LISA is approximately maximum). In the right-hand panel, eccentricities $e_{\rm LISA}$ equal to zero have been set equal to $10^{-5}$. \label{fig:eccentricity}}
}
\end{figure*}

Figure~\ref{fig:eccentricity} shows the initial eccentricity $e_{\rm in}$ of our merging BBHs (i.e. the orbital eccentricity of the binary when it becomes a BBH binary) and the eccentricity $e_{\rm LISA}$ when the orbital frequency is $f_{\rm orb}=10^{-2}$ Hz (approximately associated to the maximum sensitivity of LISA, see \citealt{amaro-seoane2017}). To estimate $e_{\rm LISA}$, we evolve the BBHs up to $f_{\rm orb}=10^{-2}$ Hz following \cite{peters1964} equations.

The initial eccentricity distribution of SC merging exchanged BBHs shows an upturn for $e_{\rm in}>0.6$ and is clearly different from both SC merging original BBHs and IBs. In contrast, when entering the LISA band, all systems (including exchanged BBHs) have significantly circularized. The minimum detectable eccentricity by LISA is  $e_{\rm LISA}\sim{}10^{-2}$, according to \cite{nishizawa2016}. Only  $\sim{}2.5$~\%{} ($\sim{}1.6$~\%{}) merging exchanged BBHs (merging BBHs in IBs) have $e_{\rm LISA}\gtrsim{}10^{-2}$. 

Our results confirm that BBHs formed by exchange have significantly larger eccentricity at formation than other BBHs (see e.g. \citealt{ziosi2014}). However, the maximum eccentricity  of young SC BBHs in the LISA band is significantly smaller than the distribution of BBHs in both globular clusters and nuclear SCs \citep{nishizawa2016,nishizawa2017,rodriguez2018}. 
We stress that the $N-$body code we use integrates resonant encounters, although it does not include an accurate treatment of post-Newtonian terms (it only reproduces the effects of 2.5 PN thanks to \citealt{peters1964} formulas).

\subsection{Delay time distribution}
\begin{figure}
  \center{
    \epsfig{figure=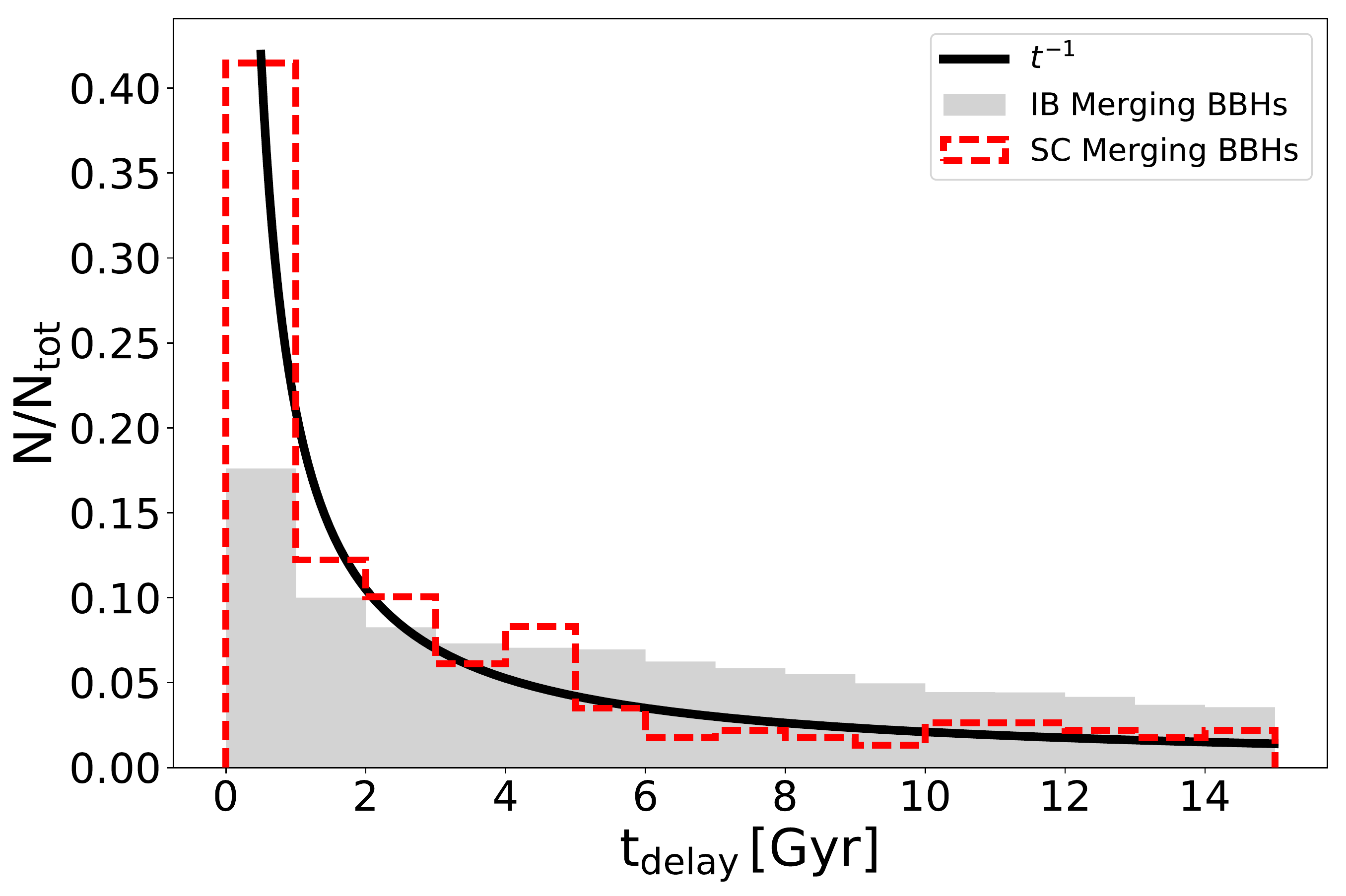,width=8.0cm} 
    \caption{Distribution of the delay timescales $t_{\rm delay}$ of merging BBHs. Gray shaded histogram: merging BBHs in IBs; red dashed line: all merging BBHs from SC simulations (exchanged BBHs plus original BBHs).} \label{fig:delaytime}}
\end{figure}
A key quantity to predict the merger rate and the properties of merging BBHs is the delay time, i.e. the time elapsed between the formation of the stellar progenitors and the merger of the two BHs.

Figure~\ref{fig:delaytime} shows that SC BBHs undergo a faster merger than BBHs in IBs. In particular, the distribution of delay times for SC merging BBHs scales as $dN/dt_{\rm delay}\propto{}t_{\rm delay}^{-1}$ (see e.g. \citealt{dominik2012}), while IB merging BBHs follow a shallower slope, with significantly less mergers in the first Gyr and a more populated tail with $t_{\rm delay}\gtrsim{}5$ Gyr. This suggests that dynamical interactions in young SCs can speed up the merger of BBHs.

\subsection{Number of mergers per unit stellar mass}
Does SC dynamics boost the merger rate, too?

The simulations presented in this paper are not sufficient to estimate the merger rate, because we considered only one stellar metallicity and we know from previous papers \citep{giacobbo2018b} that the merger rate of BBHs is very sensitive to stellar metallicity. However, we can estimate the merger efficiency, i.e. the number of mergers per unit stellar mass for a given metallicity $N_m(Z)$, integrated over the Hubble time.

As explained for the first time in \cite{giacobbo2018}, $N_m(Z)$ is the number of BBH mergers we find in the simulated sample ($N_{\rm merger}$), divided by the total initial mass of the stellar population ($M_{\ast}$): 
\begin{equation}\label{eq:nm}
N_m(Z)=\frac{N_{\mathrm{merger}}}{M_{\ast}}.
\end{equation}
For the N-body simulations, $M_{\ast}$ is just the total initial stellar mass of the simulated SCs, because the simulated SCs already include star masses ranging between 0.1 and 150 M$_\odot$ and because we use a quite realistic original binary fraction ($f_{\rm bin}=0.4$). In contrast, in the IB simulations, $M_{\ast}=M_{\ast{},sim}/(f_{\rm bin}\,{}f_{\rm corr})$, where $M_{\ast{},sim}$ is the total initial stellar mass of the simulated IBs, $f_{\rm bin}=0.4$ accounts for the fact that we are simulating only binaries and not single stars, and $f_{\rm corr}$ accounts for the missing low-mass stars between 0.1 and 5 M$_\odot$ (see Table~\ref{tab:table1}).

The values of $N_m(Z=0.002)$ for the four simulation sets presented in this paper are shown in Table~\ref{tab:table5}. The merger efficiency of the four simulation sets are remarkably similar. This implies that dynamics in young SCs does not affect the merger rate significantly.

Indeed, dynamics changes the properties of merging binaries (leading to the formation of more massive BBHs), but does not change their number significantly. The number of merging BBHs which form by exchange or harden by dynamical interactions appears to be compensated by the suppression of light merging BBHs (see Figure~\ref{fig:totalmasses}). In a forthcoming study (Di Carlo et al., in prep.) we will check whether this result depends on stellar metallicity or on other properties of the simulated SCs.

\begin{table}
\begin{center}
\caption{ Number of mergers per unit stellar mass. \label{tab:table5}} \leavevmode
\begin{tabular}[!h]{cc}
\hline
Set & $N_m(Z=0.002)$\\
    & [M$_\odot{}^{-1}$]\\
\hline
HF & $1.75\times{}10^{-5}$\\
LF & $1.61\times{}10^{-5}$\\
SC & $1.68\times{}10^{-5}$\\
IB & $1.7\times{}10^{-5}$\\
\hline
\end{tabular}
\end{center}
\begin{flushleft}
  \footnotesize{Column~1: Name of the simulation set; column~2: $N_m(Z=0.002)$ as defined in equation~\ref{eq:nm}.}
\end{flushleft}
\end{table}

\subsection{Impact of fractality}\label{sec:fractality}
So far, we have considered HF and LF simulations as a single simulation sample (SC simulations), because fractality does not significantly affect most of BBH properties (e.g. the number of merging BBHs, the merger efficiency and the mass of merging BHs, see Table~\ref{tab:table3}). Thus, we can conclude that the level of substructures does not significantly affect the merger of BBHs and can be neglected in future studies. This result is important because it removes one of the possible parameters which were thought to affect the statistics of BBHs.

However, there are a couple of intriguing differences between HF and LF simulations (see Table~\ref{tab:table3}), although these differences are only marginally significant.

\begin{figure}
  \center{
    \epsfig{figure=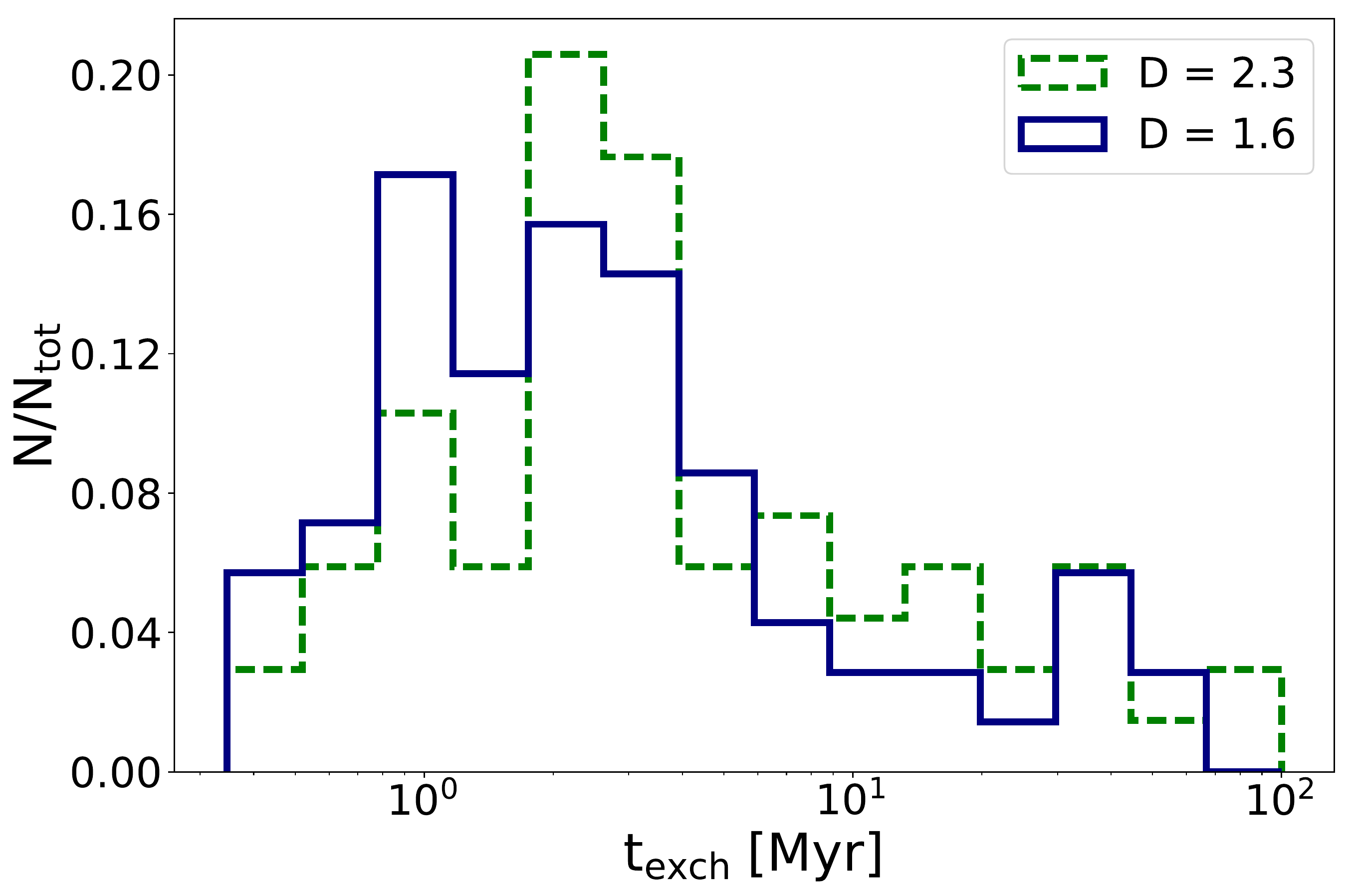,width=8.0cm} 
    \epsfig{figure=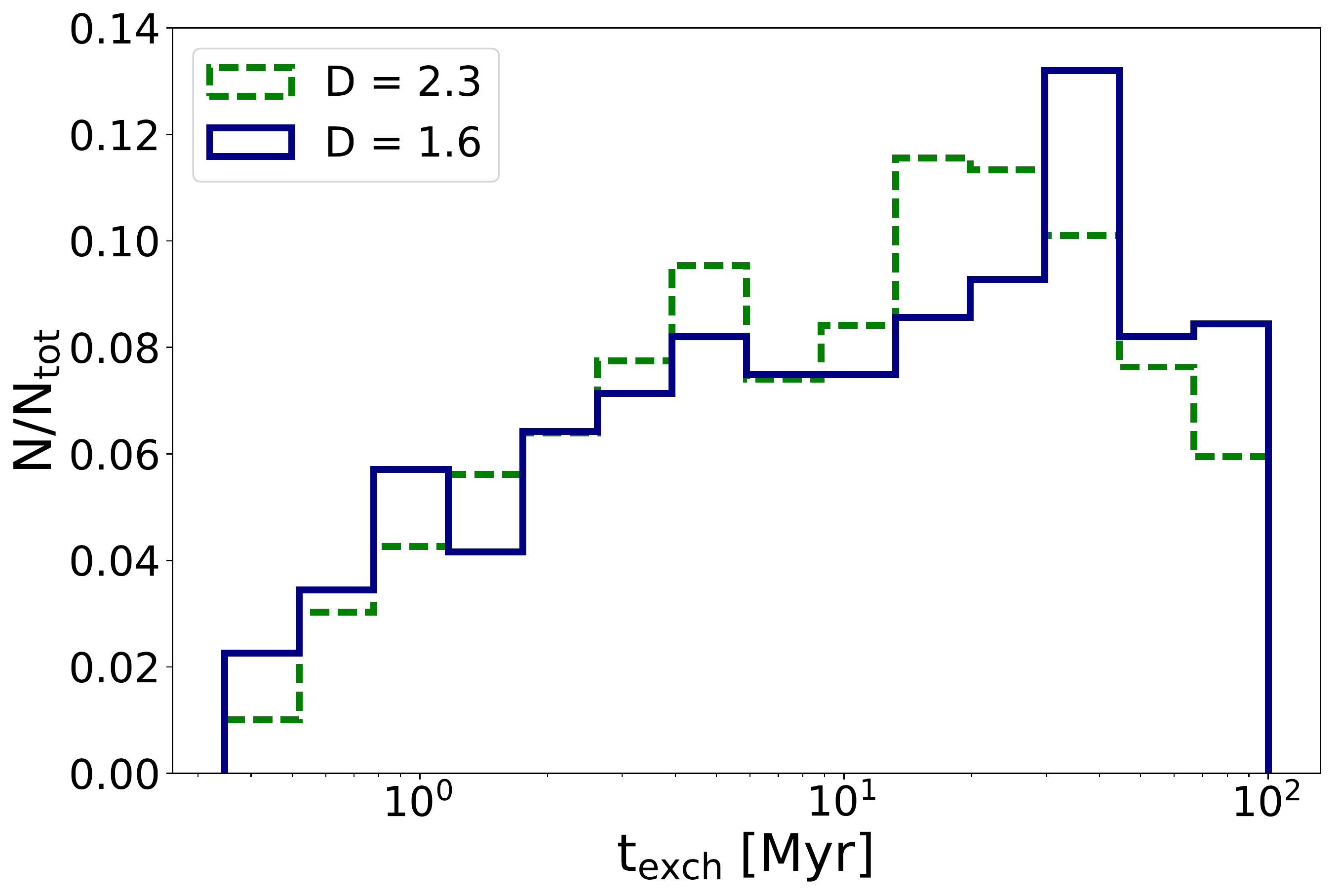,width=8.0cm} 
    \caption{Formation times of merging exchanged BBHs (top panel) and of all exchanged BBHs (bottom). The formation time is defined as the time interval between the beginning of the simulation and the moment in which the binary forms. We show the distributions for fractal dimensions $D=2.3$ (green dashed) and for $D=1.6$ (blue solid). Each distribution is normalized to its total number of elements.\label{fig:formtimes}}
}
\end{figure}

First, the HF clusters tend to produce merging BBHs with larger mass ratios than the LF clusters (median values of $q=0.9$ and $q=0.8$ in HF and LF simulations, respectively, see Table~\ref{tab:table3}). Secondly, the median time of dynamical exchange for merging exchanged BBHs ($t_{\rm exch}$, i.e. the time when a dynamical exchange leads to the formation of the binary which then becomes a merging BBH) is $t_{\rm exch}=1.9$ Myr and 2.8 Myr for HF and LF clusters, respectively (see also Figure~\ref{fig:formtimes}).

Although only marginally significant, these differences are consistent with our expectations. In fact, the more fractal (i.e. sub-structured) a SC is, the shorter is its two-body relaxation timescale $t_{\rm rlx}$ 
(see e.g. \citealt{fujii2014}). A shorter two-body relaxation timescale implies also a shorter core-collapse timescale. Core collapse is the moment of the life of a SC when three-body encounters and exchanges are more likely to occur, because of the boost in the central density. Thus, we expect dynamical exchanges (leading to the formation of merging BBHs) to occur earlier in HF simulations than in LF simulations.

Moreover, if $t_{\rm exch}<3$ Myr, the exchange occurs before the binary has turned into a BBH, because the lifetime of the most massive stars is $\gtrsim{}3$ Myr. If the binary which forms after an exchange is still composed of two non-degenerate stars (or a star and a BH), it may still undergo mass transfer and common envelope after the exchange. This mass transfer or common envelope is expected to ``redistribute'' the mass between the two members of the binary, leading to the formation of an almost equal mass BBH. In contrast, if $t_{\rm exch}\gg{}3$ Myr, the outcome of the exchange is already a BBH, which cannot undergo mass transfer and whose mass ratio can be small. For this reason, the fact that $t_{\rm exch}$ is significantly shorter in HF clusters than in LF clusters implies that the median $q$ of the former is larger than the median $q$ of the latter, because exchanged binaries in HF clusters have more chances to undergo common envelope than exchanged binaries in LF clusters.

\section{Discussion}
We have shown that young SC dynamics crucially affects the main properties (mass, eccentricity and time delay) of merging BBHs, but how common is the dynamical formation of BBHs in young SCs?

Star formation (and especially massive star formation) is expected to occur mostly in young SCs and associations \citep{lada2003,portegieszwart2010}. 
This implies that most BBH progenitors were born in SCs or OB associations, and a significant fraction of them underwent dynamical interactions before being ejected from their parent SC, or before the SC dissolved in the galactic tidal field. Is the time spent by a BBH (or BBH progenitor) in the parent SC sufficient to significantly affect the properties of merging BBHs?

The crucial point is to understand when merging exchanged BBHs form in our simulations. 
In fact, our simulations include a static tidal field, assuming that the simulated SCs are on circular orbits approximately at the location of the Sun. This approach tends to overestimate the lifetime of SCs, because we do not account for orbits closer to the galactic center and, most importantly, for massive perturbers (e.g. molecular clouds), which might accelerate the disruption of the parent SC \citep{gieles2006}.

Figure~\ref{fig:formtimes} shows the time of the dynamical exchange for merging exchanged BBHs (top) and for all exchanged BBHs (bottom). While most exchanged BBHs form at late stages in the evolution of their parent SC, the majority of merging exchanged BBHs form in the first $\sim{}10$ Myr. This guarantees that the majority of merging exchanged BBHs would have formed even if the tidal field was locally stronger.

The difference between top and bottom panel of Figure~\ref{fig:formtimes} can be explained straightforwardly with the nature of the exchanged binaries: only $\sim{}13$ \% of merging exchanged BBHs originate from a dynamical exchange which leads to the immediate formation of a BBH, while the dynamical exchange produces a new star--star binary (i.e. a binary system composed of two stars) and a BH--star binary in the $\sim{}77$ \% and $10$ \% of systems, respectively. In contrast, $\sim{}59$ \% of all exchanged BBHs form from dynamical exchanges which lead to the formation of a BBH directly, while 37\% (4\%) exchanges produce star-star (BH--star) binaries. This implies that binaries which undergo an exchange before both stars have turned into BHs are more likely to merge within a Hubble time. The reason is that exchanged star--star binaries and BH--star binaries can still undergo common envelope episodes, which shrink their orbital separation significantly and favour their merger within a Hubble time. Thus, in young SCs, stellar binary evolution and dynamics strictly cooperate to the formation of merging exchanged BBHs.

In contrast, double degenerate binaries can shrink only by three-body encounters and by gravitational wave decay, which are not as efficient as common envelope. Thus, most exchanged BBHs which form at late times cannot merge within a Hubble time. In this regard, young SCs are quite different from globular clusters \citep{rodriguez2015,rodriguez2016,askar2017}: a large fraction of exchanged BBHs in globular clusters remain inside the parent SCs for many Gyr, undergoing several exchanges and shrinking efficiently by dynamical hardening, while most exchanged binaries in young SCs undergo a single exchange and harden for a short time span.

A further crucial question is whether there are unique signatures of merging exchanged BBHs, which can be chased by gravitational-wave detectors. The masses of merging BBHs formed via  SC and IB simulations are remarkably different in our simulations. However, lower progenitor metallicity can induce the formation of more massive merging BBHs in IBs (e.g. \citealt{giacobbo2018b}). Thus, the effects of metallicity and dynamics are somewhat degenerate.

On the other hand, the most straightforward smoking gun of dynamical evolution is the formation of merging BHs in the mass range forbidden by (pulsational) pair instability ($\approx{}60-120$ M$_\odot$, \citealt{spera2017}). In our dynamical simulations, only 5 out of 229 merging BBHs ($\sim{}2$ \% of all merging BBHs) fall in this forbidden mass range. Currently, no gravitational-wave events lie in this mass range, but the detection of one such BBHs would be a strong support for the dynamical formation channel. 
In a forthcoming study, we will generalize our results to different stellar metallicities.

\section{Summary}
We have investigated the formation of BBHs in young star clusters (SCs) and open clusters. These SCs represent the bulk of star formation in galaxies. For our study, we have integrated the new population synthesis code \textsc{MOBSE}, which implements up-to-date stellar winds, binary evolution and supernova models \citep{mapelli2017,giacobbo2018,giacobbo2018b}, with the direct summation N-Body code \textsc{NBODY6++GPU}, which allows us to account for close encounters and dynamical exchanges \citep{wang2015}.

We find that dynamics significantly affects the properties of merging BBHs: dynamical exchanges favour the formation and the merger of massive BBHs with total mass up to $m_{\rm tot}\sim{}120$ M$_\odot$ and with mass ratio ranging from  $q\sim{}1$ to $q\sim{}0.1$ (although large mass ratios are more likely).

For comparison, merging BBHs evolved in isolated binaries (run with the same population synthesis code) have maximum total mass $m_{\rm tot}\lesssim{}70$ M$_\odot$ and a significantly stronger preference for large mass ratios.

The masses of merging BBHs in our simulations are consistent with the first 10 BBHs observed by the LIGO-Virgo collaboration \citep{abbottO2,abbottO2popandrate}. At the simulated metallicity ($Z=0.002$), the BH mass of GW170729 ($m_1=50.6^{+16.6}_{-10.2}$, $m_2=34.4^{+9.1}_{-10.1}$ M$_\odot$, \citealt{abbottO2}) can be matched only by dynamically formed BBHs, within the 90\% credible interval.

On the other hand, young SC dynamics does not affect the merger rate. We find almost the same merger efficiency ($N_m\sim{}1.7\times{}10^{-5}$ M$_\odot^{-1}$) in SC and in IB simulations. The formation of massive merging BBHs by dynamical exchanges is compensated by the loss of light merging BBHs, which are softened or ionized by interactions.

Almost all BBHs in our simulations merge after they were ejected from the SC or after the SC dissolved and became field.

Dynamics tends to speed up the merger of BBHs: the delay time between the formation of the stellar progenitors and the merger of the BBH scales approximately as $dN/dt_{\rm delay}\propto{}t_{\rm delay}^{-1}$ for SC merging BBHs, while the trend is much shallower for IB merging BBHs.


Finally, about 2~\% of merging BHs formed in young SCs have mass $\gtrsim{}60$ M$_\odot$, lying in the ``forbidden'' region by (pulsational) pair instability ($m_{\rm BH}\sim{}60-120$ M$_\odot$, e.g. \citealt{spera2017}). These BHs form by the merger of two or more stars. According to our prescriptions, merged metal poor stars with a Helium core smaller than $\sim{}60$ M$_\odot$ might retain a large hydrogen envelope: they avoid the pair instability window and might collapse to  massive BHs. In the field, such massive BHs remain single, while in young SCs they can acquire companions dynamically and merge by gravitational-wave emission. Thus, observing a rare merging BH  with mass $\gtrsim{}60$ M$_\odot$ would be a strong signature of dynamical formation.

\section*{Acknowledgments}
We thank all the members of the ForDyS group for their useful comments.
UNDC acknowledges financial support from Universit\`a degli studi dell'Insubria through a Cycle 33rd PhD grant.
MM acknowledges financial support by the European Research Council for the ERC Consolidator grant DEMOBLACK, under contract no. 770017. MP acknowledges financial support from the European Union’s Horizon 2020 research and innovation programme under the Marie Sklodowska-Curie grant agreement No. 664931. MS acknowledges funding from the European Union's Horizon 2020 research and innovation programme under the Marie-Sklodowska-Curie grant agreement No. 794393. LW acknowledges financial support from Alexander von Humboldt Foundation. This work benefited from support by the International Space Science Institute (ISSI), Bern, Switzerland,  through its International Team programme ref. no. 393  {\it The Evolution of Rich Stellar Populations \& BH Binaries} (2017-18).
 

\bibliography{./bibliography}
\end{document}